\documentclass[%
 twocolumn,
 superscriptaddress,
nofootinbib,
 amsmath,amssymb,
 aps,
 prl,
]{revtex4-2}

\usepackage{chemformula} 
\usepackage[T1]{fontenc} 
\usepackage[utf8]{inputenc}
\usepackage{xcolor}
\usepackage{xcolor, soul}
\usepackage[colorinlistoftodos]{todonotes}
\usepackage{colortbl} 
\usepackage{amsmath}
\usepackage{graphicx}

\usepackage{xr}
\makeatletter

\newcommand*{\addFileDependency}[1]{
\typeout{(#1)}
\@addtofilelist{#1}
\IfFileExists{#1}{}{\typeout{No file #1.}}
}\makeatother

\newcommand*{\myexternaldocument}[1]{%
\externaldocument{#1}%
\addFileDependency{#1.tex}%
\addFileDependency{#1.aux}%
}

\myexternaldocument{FeBr2_SI_arxiv}

\newcommand{\FeBr}{FeBr$_{2}$}
\newcommand{\FeBrx}{FeBr$_\mathrm{x}$}
\newcommand{\CoBr}{CoBr$_{2}$}
\newcommand{\vFeBr}{V$_\mathrm{Br}-$FeBr$_{2}$}
\newcommand{\vCoBr}{V$_\mathrm{Br}-$CoBr$_{2}$}
\newcommand{\vvFeBr}{2V$_\mathrm{Br}-$FeBr$_{2}$}

\newcommand{\moire}{Moir\'{e} }

\begin{document}

\title{Intrinsically patterned two-dimensional transition metal halides}

\author{Feifei Xiang}
\thanks{These authors contributed equally\\}
\affiliation
{Department of Physics, Friedrich-Alexander-Universität Erlangen-Nürnberg, Erlangen, Germany}
\author{Neeta Bisht}
\thanks{These authors contributed equally\\}
\affiliation
{Department of Chemistry and Pharmacy, Chair of Theoretical Chemistry, Friedrich-Alexander-Universität Erlangen-Nürnberg, Erlangen, Germany}
\author{Binbin Da}
\thanks{These authors contributed equally\\}
\affiliation
{Department of Physics, Friedrich-Alexander-Universität Erlangen-Nürnberg, Erlangen, Germany}
\author{Mohammed S. G. Mohammed}
\affiliation
{Department of Physics, Friedrich-Alexander-Universität Erlangen-Nürnberg, Erlangen, Germany}
\author{Christian Nei\ss}
\affiliation
{Department of Chemistry and Pharmacy, Chair of Theoretical Chemistry, Friedrich-Alexander-Universität Erlangen-Nürnberg, Erlangen, Germany}
\author{Andreas Görling}
\email{andreas.goerling@fau.de}
\affiliation
{Department of Chemistry and Pharmacy, Chair of Theoretical Chemistry, Friedrich-Alexander-Universität Erlangen-Nürnberg, Erlangen, Germany}
\author{Sabine Maier}
\email{sabine.maier@fau.de}
\affiliation
{Department of Physics, Friedrich-Alexander-Universität Erlangen-Nürnberg, Erlangen, Germany}

\date{\today}

\begin{abstract} 
Patterning and defect engineering are key methods to tune 2D materials' properties. However, generating 2D periodic patterns of point defects in 2D materials has been elusive until now, despite the well-established methods for creating isolated point defects and defect lines. Herein, we report on intrinsically patterned 2D transition metal dihalides on metal surfaces featuring periodic halogen vacancies that result in alternating coordination of the transition metal atoms throughout the film. Using low-temperature scanning probe microscopy and low-energy electron diffraction, we identified the structural properties of patterned \FeBr \ and \CoBr \ monolayers grown epitaxially on Au(111). Density-functional theory reveals that the Br-vacancies are facilitated by low formation energies and accompanied by a lateral softening of the layers leading to a significant reduction of the lattice mismatch to the underlying Au(111). We demonstrate that interfacial epitaxial strain engineering presents a versatile strategy for controlled patterning in 2D. In particular, patterning 2D magnets provides new pathways to create unconventional spin textures with non-collinear spin.
\end{abstract}

\maketitle


\section{Introduction}
Defect engineering and polymorphisms are two widely used concepts to create novel architectures and introduce new functionalities into two-dimensional (2D) materials. These concepts have been widely studied in van der Waals (vdW) materials like transition metal dichalcogenides (TMDCs), with much attention given to polymorphs, point defects, and line defects. However, the controlled assembly of 2D periodic patterns of point defects has remained elusive.\cite{Feng2016, Lin2017} Vacancy lattices are particularly interesting for the selective functionalization of 2D materials with atoms and molecules as well as tuning their electronic properties. Instead, patterning in 2D was focused intensely on \moire patterns using vdW-materials heterostacks.\cite{He2021} \\

We demonstrate here the controlled 2D patterning in single-layer transition metal halides (TMH). TMH gained significant interest since the recent discovery of intrinsic ferromagnetism in 2D vdW-materials.\cite{Gong2017,Huang2017} Hence, the electronic and magnetic properties of first-row 2D transition metal trihalides MX$_3$ (M = V, Cr, Mn, Fe, Co, Ni; X = Cl, Br, I) came into the focus of first-principles calculations and experimental studies.\cite{Wang2022} In contrast, only a limited number of experimental surface-science studies are available for 2D transition metal dihalides (TMDs) so far that provide atomic scale insights on the structure, growth, and defects in real space.\cite{Jiang2023,Bikaljevic2021,Zhou20,Cai2021,Cai2020,Liu2020,Zhou2022} This might be related to challenges in their preparation: On the one hand, some materials decompose  during the growth by molecular beam epitaxy (MBE) in ultra-high vacuum (UHV) associated with a loss of halogens.\cite{Bikaljevic2021} On the other hand, some \textit{ex-situ} prepared and exfoliated layers suffer from limited environmental stability in ambient conditions.\cite{Shcherbakov18,Mastrippolito2021}\\

The study of polymorphism and intermediate stoichiometries (different from MX$_{2}$ and MX$_{3}$) in 2D transition metal halides is just beginning.\cite{Bergeron2021,Han2023} Analogous to the structure of TMDC, TMD  monolayers can adopt both trigonal prismatic (1H) or octahedral coordination (1T) of the metal cation, see  Fig~S1.\cite{Gosh21,Kulish2017} However, 1T stacking is the energetically most favorable stacking for most TMHs. In conclusion, strategies for stabilizing meta-stable polymorphs and periodic patterning for 2D materials, in general, require further research. In this respect, interface engineering using lattice misfits is a powerful and promising tool, as will be outlined. \\

\begin{figure*}[t!]
 \includegraphics[width=0.8\paperwidth]{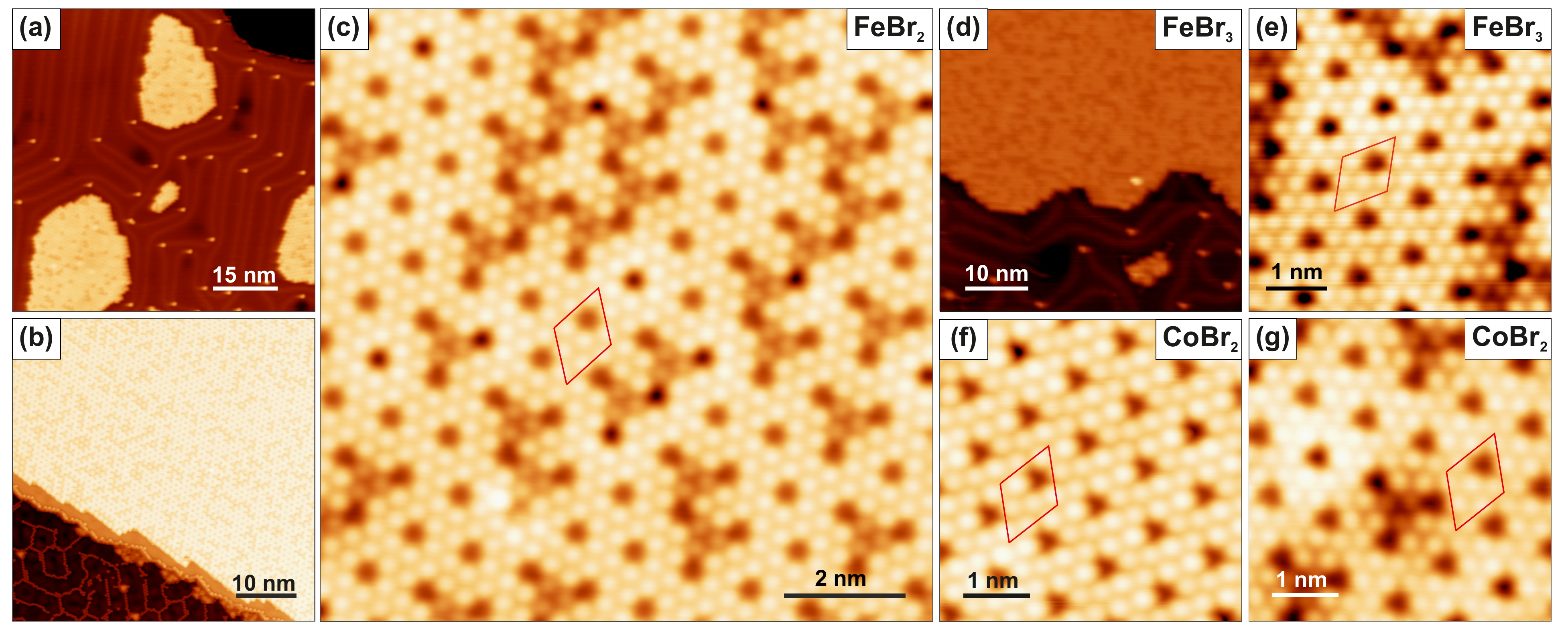}
 \caption{\textbf{STM of \FeBr \ and \CoBr \ with a Br vacancy lattice on Au(111).} (a-c) STM images of \vFeBr \ at (low and high) submonolayer coverage upon deposition of  \FeBr \ powder \ on Au(111) kept at 450~K.  (d-e) STM images of \vFeBr \  after deposition of 0.4 ML FeBr$_{3}$ powder on Au(111) kept at 390~K.  (f-g) STM images of \vCoBr \ after submonolayer deposition of \CoBr \ powder on Au(111) held at 390~K. While (f) shows perfect \vCoBr, in (g) typical triangular-shaped defects are seen similar to \vFeBr. STM parameters: (a) $I=100$~pA, $U=-1$~V; (b) $I=100$~pA, $U=100$~mV; (c) $I=800$~pA, $U=50$~mV; (d) $I=100$~pA, $U=-500$~mV;
 (e) $I=100$~pA, $U=-50$~mV; (f) $I=100$~pA, $U=5$~mV; (g) $I=100$~pA, $U=10$~mV.}
 \label{fig:structuresyntesis_exp}
\end{figure*}

Here, we report on the growth and characterization of intrinsically patterned single-layer iron bromide (\FeBr) \ and cobalt bromide (\CoBr) \ on Au(111). Interestingly, the periodic arrangement of halogen vacancies results in an alternating coordination (6-fold and 5-fold) of the transition metal atoms across the film.  The decomposition during thermal evaporation of \FeBr \ and \CoBr \ powder facilitates the formation of halogen-vacancy lattices in the respective 2D layers. The Br-vacancies are realized due to their low formation energies and stabilized by a significant reduction of the misfit strain at the TMD-Au interface. Detailed structural characterization using low-temperature scanning tunneling microscopy (STM), non-contact atomic force microscopy (nc-AFM), low-energy electron diffraction (LEED), and density-functional theory (DFT) provides comprehensive insights into the interfacial properties of intrinsically patterned \FeBr \ and \CoBr \ monolayers on Au(111). The alternating coordination number of the transition metal atoms throughout the films, opens the way to intriguing magnetic and electronic properties, making the films interesting candidates for applications in spintronics.\cite{Mak2019,Feng2018} In particular, patterning \FeBr \ and \CoBr, which are predicted to be 2D ferromagnets,\cite{Botana19,Kulish2017} provides new pathways to create unconventional spin textures with non-collinear spins.\\

\section{Results}

\subsubsection{\textbf{Synthesis and structure of intrinsically patterned 2D \FeBr \ and \CoBr}}

Fig.~\ref{fig:structuresyntesis_exp}a-c show STM images illustrating the typical morphology of single layer \FeBrx \ films grown in UHV by thermal evaporation of anhydrous \FeBr \  and FeBr$_{3}$ powder on Au(111) kept at 450~K and 390~K, respectively. It is remarkable that the structure of both powders is almost identical, providing direct evidence for non-stoichiometric sublimation. The atomic-resolution STM images reveal a hexagonal lattice with a period of $392\pm 30$~pm, similar to the unit cell of a \FeBr \ single crystal with 377.2 pm,~\cite{Wilkinson59} which resembles the Br atoms in the top TMD halide layer. In addition to the atomically resolved Br structure, the STM topographies exhibit a well-ordered hexagonal superstructure of dark depressions ($60-110$~pm in depth in STM) highlighted by the red unit cells. The superstructure  has a size of $10.37\pm0.3$~\AA \ and is mostly defect-free over the entire layers. We note the superstructure also formed at room temperature preparations.  In addition to the regular superstructure, triangular-shaped irregularities in the top Br lattice are seen, which we attribute to Br-bottom defects and which will be discussed in the next section. \\ 

\begin{figure*}[t!]
 \includegraphics[width=0.8\paperwidth]{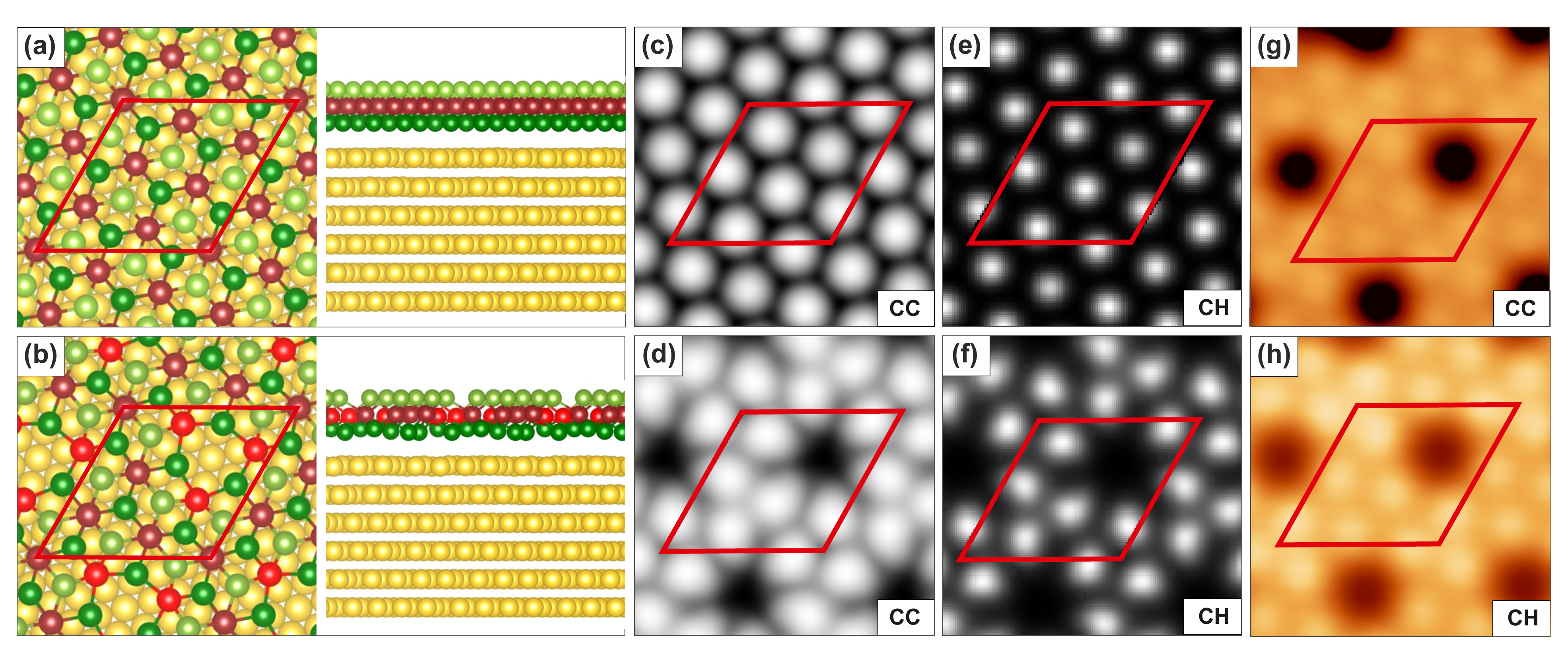}
 \caption{\textbf{Structure of pristine and patterned \FeBr \ films on Au(111).} (a-b) DFT-optimized structural model of pristine \FeBr \ and \vFeBr \ on Au(111) with corresponding calculated constant-height (CH)  and constant-current (CC) STM images at 500~mV in (c-f). (g-h) Experimental constant-current (420mV, 200~pA) and constant-height (420~mV) STM images of \vFeBr \ on Au(111). Color code: yellow, Au; light green, top Br; dark green, bottom Br; dark red, 6-fold coordinated Fe; red, 5-fold coordinated Fe.  }
 \label{fig:structuresyntesis_theory}
\end{figure*}

The superstructure can be attributed to either an ordered vacancy lattice or a \moire pattern caused by the rotation of the \FeBr \ lattice  with respect to the Au(111) lattice. Generating periodically arranged point defects in 2D materials with this high quality is an extremely challenging task, however, and remains elusive to date.\cite{Feng2016,Lin2017} Nevertheless, we assign the occurrence of the superstructure to a periodic Br vacancy lattice for the following reasons: (i) Apart from the non-stoichiometric deposition of \FeBr \ and FeBr$_{3}$ resulting in the same 2D-TMD structures,  we frequently observe  atomic chains on Au(111) forming a mesh (see Fig.~\ref{fig:structuresyntesis_exp}b), which we assign to residual Br atoms coexisting to the 2D-TMD islands. Both corroborate a thermal decomposition of the powders during sublimation, leading to Br-deficient 2D-TMD. (ii) The topographic contrast of the superstructure is mostly bias-independent over a large voltage range (-3V to 2 V, see Fig.~S2) and also clearly seen in constant-height mode STM and nc-AFM images, which rules out that the superstructure is related to an electronic effect, for instance, an electronic \moire pattern arising from the rotated TMD with respect to the Au lattice, see discussion about defects below. (iii) Only one atom in the unit cell is recessed from the surface; there is no periodic modulation observed, common for \moire patterns. \\

Next, we tried to reproduce the experimental STM images using DFT calculations at the PBE+D3 level using VASP to  corroborate the conclusion that the observed superstructures are vacancy lattices in the top halide layer. The unit cell of the pristine \FeBr \ structure contains 21 atoms (7 Fe and 14 Br), matching the unit cell of the superstructure observed experimentally. Initially, the pristine and defective 1T-\FeBr \ layers were relaxed in the gas phase and subsequently deposited onto the surface using a commensurate unit cell size. As expected, the periodic depressions are not reproduced in calculated STM images of a DFT-optimized pristine 1T-\FeBr \ layer on Au(111), see  Fig.~\ref{fig:structuresyntesis_theory}.  In fact, the calculated STM image of the pristine \FeBr \ layer on Au(111) shows only the atomic corrugation of the Br atoms in the top layer, not even a \moire pattern, see also Fig~S3. In order to identify the most likely point defect causing the periodic depressions in the experimental images, we calculated \FeBr-films with periodic Br-vacancies in the top layer, as the depressions coincide with Br-sites. The simulated constant-current and constant-height STM images show a depression at the Br-vacancy site in good agreement with experimental images, as depicted in Fig. 2c-f. Importantly, introducing the Br vacancies in the top layer leads to a periodic alternation of 5-fold and 6-fold coordinated Fe atoms throughout the films (highlighted by light and dark-colored Fe atoms in Fig.~\ref{fig:structuresyntesis_theory}a-b, which may lead to intriguing magnetic and electronic properties. Thus, the excellent match between the experimental findings and the simulated STM calculations supports that a Br-vacancy lattice in the top Br-layer of the TMD is formed. We refer to the Br-vacancy lattices as  V$_\mathrm{Br}-$FeBr$_{2}$ \ and V$_\mathrm{Br}-$CoBr$_{2}$, in the following.\\

Interestingly, the formation of a Br vacancy lattice in the Br top layer is not unique to \FeBr, but was also observed for submonolayers of \CoBr \ deposited on the Au(111) kept at 390~K, see Fig.~\ref{fig:structuresyntesis_exp}f. We note the similarity in lattice parameters between the two TMDs, which may imply that the structures could be stabilized by the mismatch to the Au surface.\cite{Kulish2017}  However, the structure is different for NiBr$_{2}$, likely because for NiBr$_{2}$ thermal decomposition is negligible.\cite{Bikaljevic2021} In the following, we mainly focus on \FeBr.\\

\begin{figure*}[t!]
\includegraphics[width=0.65\paperwidth]{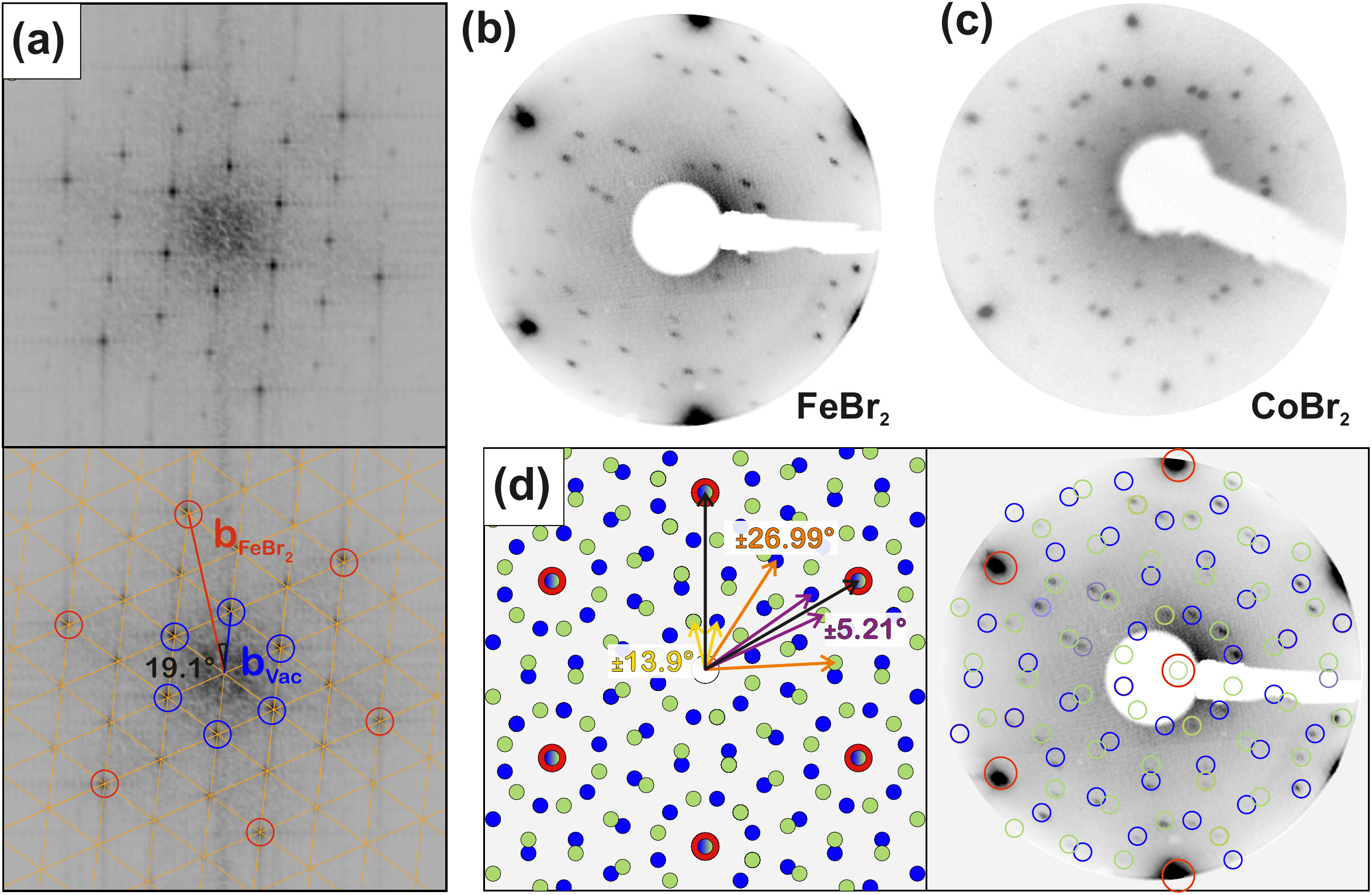}
\caption{\textbf{Superstructure of \vFeBr \ and \vCoBr \ determined by STM and LEED.} (a) Fast Fourier transform of an atomically resolved STM image of a \vFeBr \ monolayer.  The orange grid confirms that the Br vacancy lattice (blue) is commensurate to the \FeBr \ lattice (red), adopting a $(\sqrt{7}\times\sqrt{7})\mathrm{R}19.1^{\circ}$ structure. (b-c) LEED images of submonolayer ($\approx0.4$~ML) \vFeBr \ (47.1~eV) and \vCoBr \ (64.9~eV), revealing similar superstructure patterns. \vFeBr \ was deposited at 450~K and \vCoBr \ at 390~K, respectively. We note the LEED of \vCoBr \ was performed with an MCP LEED, resulting in some distortions.\cite{Sojka2013}
(d) Simulated LEED pattern of the superstructure described by the matrix notation  $\big(\begin{smallmatrix}
  4 & 1\\
  -1 & 3
\end{smallmatrix}\big)$ and 
$\big(\begin{smallmatrix}
  4 & 3\\
  -3 & 1\end{smallmatrix}\big)$ (green and blue) and the Au substrate (red). } 
\label{fig:powder_superstructure}
\end{figure*}
   
 The Br-vacancy superstructure denotes a ($\sqrt{7}\times\sqrt{7})\mathrm{R}19.1^{\circ}$ structure with respect to the \FeBr \ lattice, as seen by the Fast Fourier transform of an atomically resolved STM image of a \vFeBr \ layer in Fig.~\ref{fig:powder_superstructure}a. Furthermore, the unit cell with $L=\sqrt{13}a_\mathrm{Au}$ is commensurate to the Au lattice as proven by LEED, Fig.~\ref{fig:powder_superstructure}b-d. We observe a $\pm13.9^{\circ}$ rotation of the superstructure toward the high symmetry axes of Au lattice, as well as a  $\pm 5.21^{\circ}$ and $\pm 26.99^{\circ}$ rotation of the \FeBr \ lattice toward the Au lattice (Fig.~S7/Fig.~S8). Hence, there are four distinct domains with $L=\sqrt{13}a_\mathrm{Au}$, see Tab.~S1, that result in a total of 24 domains, including the ones rotated by multiples of 60$^{\circ}$ with respect to $L$,  see Fig.~S6. In accordance, the simulated LEED pattern leads to two unique subpatterns shown in blue and green in (Fig.~\ref{fig:powder_superstructure}d). The simulated LEED fits the experiment perfectly and also confirms that the observed \FeBr \ superstructure is commensurate with the underlying Au substrate.  Similar LEED patterns for \vFeBr \ and \vCoBr \ in Fig.~\ref{fig:powder_superstructure}b-c demonstrate that halogen vacancy lattices can be obtained for several TMDs. \\

The apparent height of the \vFeBr \  monolayers measures 1.8$\pm$0.1~\AA \  (Fig.~S10), which is significantly lower compared with the known bulk interlayer spacing of 6.23~\AA.\cite{McGuire2017,Haberecht2001} The lower apparent height may be explained by the difference in the local density of states of the \vFeBr \ and the Au probed by the STM measurements at the applied bias voltage, as also observed for TMDCs on Au(111).\cite{Tumino2019}  Our DFT calculations showed for the pristine \FeBr, \FeBr \ with Br top vacancy, and \FeBr \ with Br top and bottom vacancy  an averaged vertical Br-Br distance between the top and bottom layer of 2.79~\AA, 2.75~\AA, and 2.68~\AA, respectively. The pristine \FeBr \ has an averaged adsorption height of 2.91~\AA \ and 4.33~\AA \ for the bottom Br and the Fe atoms above the surface. The herringbone reconstruction of the Au(111) surface is lifted by the \vFeBr \ islands, leading to an irregular orientation of the solution lines around the \vFeBr \ islands.\\

\begin{figure*}[t!]
\includegraphics[width=0.8\paperwidth]{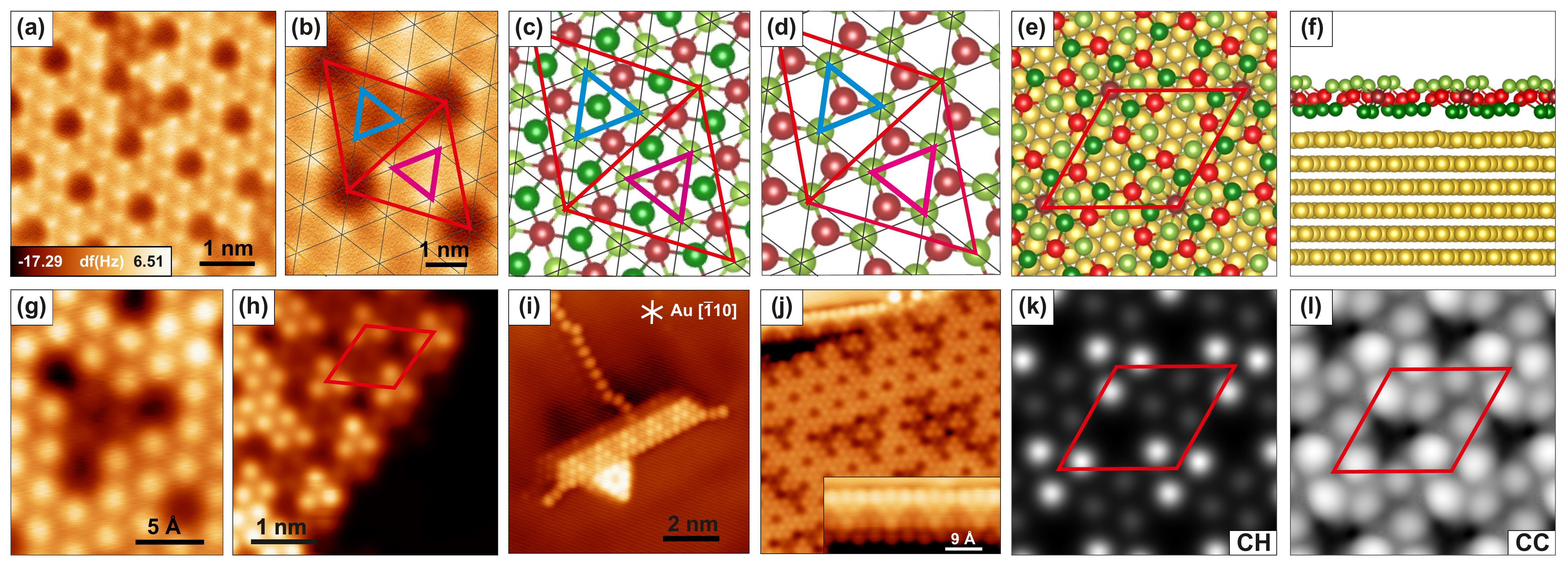}
\caption{\textbf{Defects in \FeBr \ identified by STM, nc-AFM, and DFT.} (a-b) Constant-height nc-AFM images of \FeBr \ islands with a triangular-shaped defect. The overlaid grid in (b) represents the Br lattice in the top layer. The unit cell can be divided into two parts. The triangular-shaped defect always occurs in the same half of the unit cell, which either leads as shown in (c) in a 1T-\FeBr \ to a Br bottom  (blue) or Fe vacancy (purple) and in (d) in a hypothetical 1H-\FeBr \ to a Fe defect (blue). (e-f) DFT-optimized 1T-\FeBr \ on Au(111) with one top and one bottom Br vacancy and corresponding simulated constant-height and constant-current STM ($U =0.5$~V) in (k-l). (g-h) Single and multiple triangular-shaped defects imaged by constant-height STM. (i-j) Along Au-steps or dislocation lines on Au(111) narrow \FeBr-ribbons are observed. These are the only structures without Br vacancies in the top layer. 
 Color code: yellow, Au; light green, top Br; dark green, bottom Br; dark red, 6-fold coordinated Fe; red, 5-fold coordinated Fe. STM/nc-AFM parameters: (a) z=80 pm with respect to set point $U =5$~mV, $I=560$~pA; (g) $U= -50$~mV, z = constant (inverted); (h) $U=400$~mV (i) $I=560$~pA, $U=50$~mV; (j) $I=400$~pA, $U=20$~mV; inset: $I=300$~pA, $U=20$~mV.}  
\label{fig:powder_defects}
\end{figure*}


\subsubsection{\textbf{Defects in single layer \FeBr \ and \CoBr}}

Next, we discuss the Br-vacancies in the top layer and other typical point defects observed in single layers \FeBr \ and \CoBr \ based on  STM and nc-AFM measurements. In the constant-height nc-AFM images of \FeBr, Fig.~\ref{fig:powder_defects}a-b, we assign the hexagonal lattice of bright features (positive frequency shifts) to the outer bromine atoms, which are close enough to the tip to generate repulsive forces, and the darker surrounding and depressions to the lower-lying metal atoms and Br-vacancy, respectively, whose larger distance from the tip resulted in attractive forces, see also Fig.~S5. The simulated nc-AFM images based on the DFT-optimized models on Au(111) using the probe particle model with a Br at the tip apex show excellent agreement with the experimental nc-AFM data, see Fig.~S4 The nc-AFM data corroborate that the periodic depressions are related to topographic effects, such as vacancies or atoms recessed into the TMD layer. Additionally, we can exclude substitutional defects since Br and Fe atoms possess similar vdW radii. They would result in similar nc-AFM images and are not expected to appear as a depression in the halides lattice. \\

The triangular-shaped defect, embraced by three Br vacancies and composed of three dimmer atoms surrounding a depression, can be observed in both constant-height STM (Fig.~\ref{fig:powder_defects}g-h) and nc-AFM images with a similar appearance in Fig.~\ref{fig:powder_defects}a-b. This type of defect most frequently occurs  and is independent of the preparation conditions of both 2D-TMDs, \FeBr\ and \CoBr. The simulated nc-AFM images have revealed that in \vFeBr\ single layers, the Br atoms are observed as bright protrusions. Hence, the overlaid lattice in Fig.~\ref{fig:powder_defects}b confirms that the periodic depressions of the superstructure and, likewise also, the three dimmer atoms of the triangular-shaped defect are centered at Br-sites within the halide top layer. Consequently, the dark depression in the center of the triangular-shaped defect is either a Fe vacancy (purple triangle in Fig.~\ref{fig:powder_defects}c) or a Br vacancy (blue triangle in Fig.~\ref{fig:powder_defects}c) in the bottom layer. We note that all the triangular-shaped defects have the same orientation, i.e. are in the same half-unit cell within the layer. Therefore, only one of the proposed defects occurs in the experiment. As the formation energy for halide vacancies is significantly lower than for transition metal or for antisite defects in TMDs,\cite{Cayhan21} we assign the dark depressions in the center of the triangular defect to vacancies in the Br-bottom layer. This allows the surrounding three top-Br to relax toward the surface, consistent with a lower apparent contrast in constant-height STM and nc-AFM images. We note the triangular-shaped defects can also form clusters near step edges, see Fig.~\ref{fig:powder_defects}h.  In conclusion, we dominantly see Br vacancies, in agreement that we observe a decomposition and loss of halogens during the growth of \FeBr, while Fe-based vacancies are very rare, see Fig.~S11. Hence we confirm that in \vFeBr, the Fe-sublayer is intact in contrast to FeBr$_{3}$. \\

Based on similar reasoning, it can be concluded that \vFeBr \ layers exhibit 1T-stacking as opposed to a 1H-stacking. If one assumes a 1H-\FeBr \ model, as shown in Fig. \ref{fig:powder_defects}d, the formation of the triangular defect would require an energetically expensive Fe vacancy, which may not result in the spontaneous formation of the triangular-shaped defects, which we see under all preparation conditions and for all three material systems, i.e. \FeBr, \CoBr, and FeBr$_{3}$. \\

Our next step is to simulate the appearance of the bottom layer Br vacancy using DFT calculations. We used the model for the pristine layer and selected the two Br vacancy sites, one in the top and one in the bottom layer, symmetrically positioned within the unit cell, such that both vacancy sites are located over the same Au- adsorption site (top-site). The simulated constant-height and constant-current STM images in Fig. \ref{fig:powder_defects}k-l show that the three Br in the top layer located around the defect has a reduced apparent height. The observation of a $\sim10$~pm lowered adsorption height of the surrounding Br over the Au surfaces in the optimized structure agrees well with the experimental STM images.  \\

We note some of the periodic Br-top-vacancies have a darker contrast ($\sim40$~pm height difference) in constant-current STM experiments, as seen, for example, in Fig.~S12. This is likely due to the adsorption of atoms into the vacancy, which can be H from the rest gas in UHV or free Br atoms that are covalently bound to the Au substrate, hence with a lower adsorption height above the surface and not coordinated to Fe. The latter would partially restore the 1:2 (Fe:Br) stoichiometry on the surface\\

\begin{figure}[t!]
\includegraphics[width=\columnwidth]{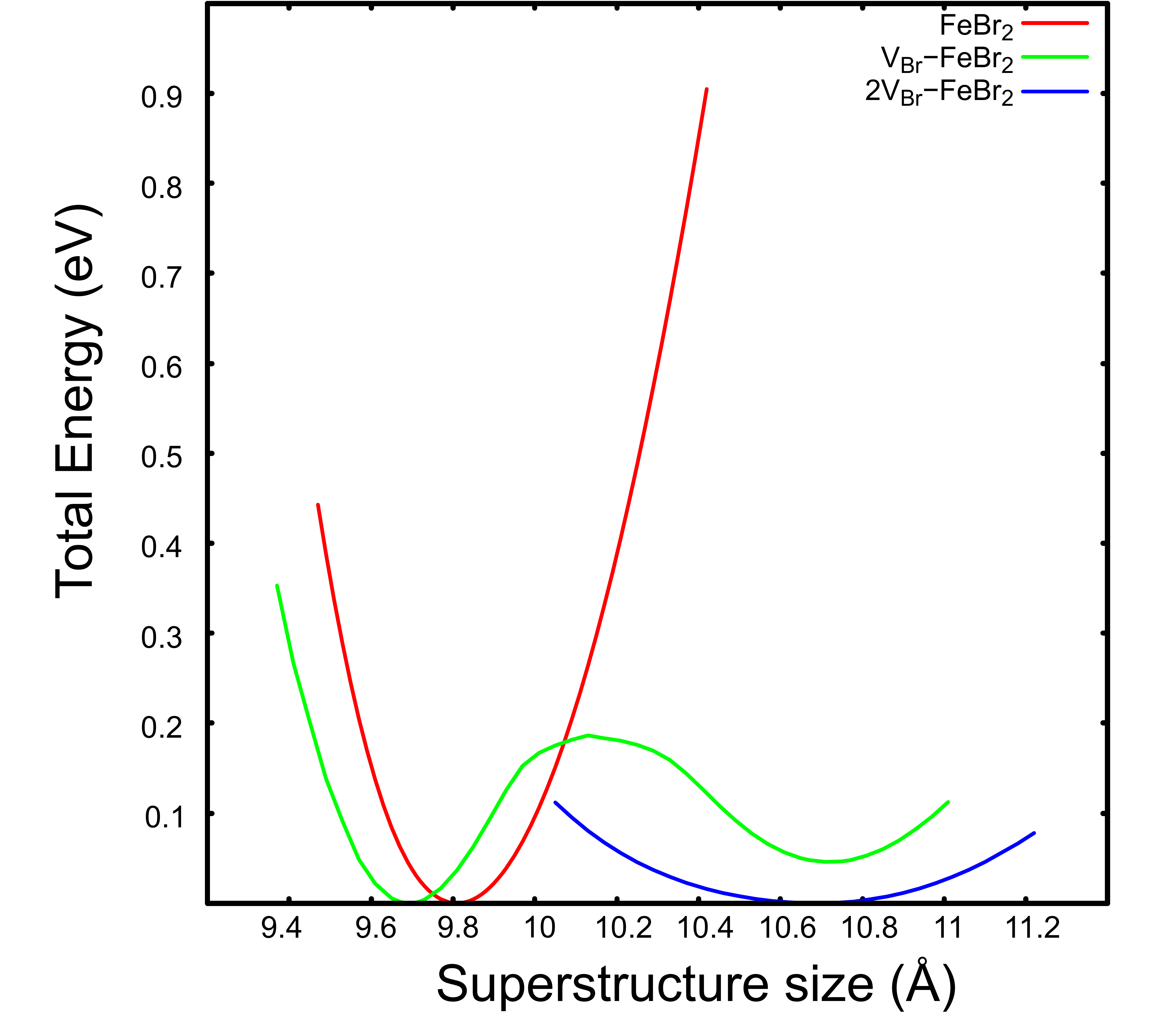}
\caption{\textbf{Growth mechanism of patterned \FeBr \ and vacancy formation based on insights from DFT and STM.} Potential energy curve for pristine \FeBr, \vFeBr, and \vvFeBr. The energy of the equilibrium state is set to 0~eV in each curve. }  
\label{fig:powder_formation}
\end{figure}

\subsubsection{\textbf{Growth mechanism of \vFeBr}}

In this section, we show how the observed bromine vacancies in \FeBr \ on Au(111) soften the monolayer and provide a possible way to reduce lattice-mismatch strain to the substrate. Importantly, the pristine \FeBr \ lattice would need to be stretched by around 4$\%$ to fit Au(111). Instead, the occurrence of periodic vacancies can lead to an expansion or compression of the lattice in TMDs,\cite{Cayhan21} e.g. the interatomic distances between transition metal cations within the layers is 3.69~\AA \ in FeBr$_{3}$ and 3.78~\AA \ in \FeBr.\cite{McGuire2017} Therefore, we investigated the softness of the pristine (\FeBr) as well as one- (\vFeBr) and two-(\vvFeBr) Br vacancies in \FeBr \ monolayers based on the energetic and structural evolution as a function of biaxial strain using DFT. As compared to the automated unit cell optimization offered by VASP, this approach has an advantage in that it simplifies the identification of multiple potential energy surfaces (PESs) that may exist due to the presence of energetically similar minima. Fig.~\ref{fig:powder_formation} shows the potential energy for freestanding monolayers as a function of superstructure size. Since the energy continuously varies without abrupt decrease, we can conclude that Fe-Br bonds are not broken, and only elastic deformation of layers is taking place. The pristine \FeBr \ and \vvFeBr \ monolayer show a single minimum at 9.81~\AA \ and 10.70~\AA, respectively.  The \vFeBr \ lattice features two local energy minima at 9.68~\AA \ and 10.71~\AA \ corresponding to the equilibrium and a meta-stable state. With an energy difference of less than 100~meV, the meta-stable state is experimentally accessible. Hence, the vacancies soften the \FeBr \ layer toward larger unit cell sizes due to incorporating 5-fold coordinated Fe atoms into the lattice. Therefore, the lattice misfit to the relaxed Au cell (10.45~\AA) is significantly reduced from the pristine to a meta-stable \vFeBr \ from 6.1\% to -2.5\% and to a \vvFeBr \ lattice to -2.4\%, respectively. \\

In order to better understand the possibility of defect formation in \FeBr \ single layers, we calculated the formation energies for Br vacancies in the top and bottom halide layers in gas phase. The calculated formation energy for a single Br vacancy in pristine \FeBr \ to obtain \vFeBr \ is 45~meV/Br, while for two Br vacancies (one in the top- and one in the bottom halide layer) to get \vvFeBr \ is  56~meV/Br. Hence, the calculated formation energies are comparable to the thermal energy at room temperature, which indicates that the formation of Br defects is energetically feasible. A low formation energy is crucial for the growth of long-range ordered and high-quality vacancy lattices. It agrees well with the experimental observations that a perfect Br-vacancy lattice in the top layer is achieved independent of film size. Only near Au-steps or dislocation lines on Au(111) narrow \FeBr \ ribbons are observed without Br-top vacancies, see Fig.~\ref{fig:powder_defects}i-j.  The formation of Br vacancies in the bottom halide layer might be diffusion-limited and hence less frequently observed, consistent with that larger islands of \vvFeBr \ are observed along layer edges. The \vFeBr \ lattices on Au(111) are stable up to around 470~K, before they desorb intact from the surface. In conclusion, we find that the growth of the periodic vacancy lattice is facilitated by strain engineering at the TMD-gold interface in combination with the low energy costs for Br vacancy formation.

\section{Discussion}
Intrinsically patterned 2D materials are crucial for the selective functionalization by the adsorption of molecules or atoms, as well as tuning their electronic and magnetic properties. While the introduction of periodic Fe vacancies into single-layer \FeBr \ leads to the formation of single-layer FeBr$_{3}$,\cite{Pere2014,Ashton2017b}  we achieved the first time the formation of periodic halogen vacancy lattices in TMDs by strain engineering. Importantly, the halogen vacancy lattices offer a way to modulate the coordination number of Fe along the 2D layer and hence alter the charge distribution, while in pristine FeBr$_{3}$ and \FeBr \ the iron coordination is constant in the entire film. We confirm previous results that 2D-\FeBr \ is ferromagnetic, see Fig.~S9.\cite{Kulish2017} The change in coordination opens pathways to exotic spin textures with non-collinear spin, which is challenging in pristine 2D-TMDs as their properties are mostly inherited from their vdW bulks and hence have a simple spin texture. Moreover, the first-time intermediate stoichiometric \FeBr \ and \CoBr \ layers (different from MX$_{2}$ and MX$_{3}$) are studied in this work, i.e. Fe$_\mathrm{7}$Br$_\mathrm{13}$ and Co$_\mathrm{7}$Br$_\mathrm{13}$. The formation of 2D materials with intrinsic vacancy patterns is likely not exclusive to \FeBr \ and \CoBr \ on Au(111). Other TMDs and eventually also TMDCs, are likely to exhibit similar patterns, provided that the formation energy of the vacancies is low and the lattice-mismatch strain is appropriately controlled in the respective system.\\

\section{Conclusions}
In conclusion, we used low-temperature STM, nc-AFM, LEED, and DFT to identify the structure of single \FeBr \ and \CoBr \ layers on Au(111). Interestingly, we observe a periodic superstructure of Br vacancies in the top halide layer on both materials. We show that the formation of the regular vacancy lattice is explained by low Br-vacancy formation energy accompanied by an increased softness of the TMD layers and a significant decrease in the lattice-mismatch strain. Despite the large number of transition metal halide and chalcogenide materials available, their periodic patterning in 2D by creating point defects has been elusive so far. The versatile strategy for long-range ordered patterning is crucial for selective functionalization with molecular and atomic species, as well as tuning their electronic and magnetic properties. In particular, as \FeBr \ and \CoBr \ are 2D ferromagnets, this provides new pathways to create unconventional spin textures with non-collinear spin.
\\

\section{Experimental Methods}

\textbf{STM and nc-AFM}. The STM and nc-AFM measurements were carried out at T = 4.8~K in an ultrahigh vacuum (UHV) system with a base pressure lower than 3$\times 10^{-10}$~mbar. A commercial STM/nc-AFM (Scienta-Omicron GmbH) equipped with a Nanonis control unit (SPECS GmbH) was used in this work. A qPlus tuning fork sensor\cite{Giessibl2000} ($k\approx 1800$~Nm$^{-1}$, $f_{0}\approx 24.95$~kHz, $Q\approx 13900$) with a chemically etched tungsten tip was used to acquire  most STM and nc-AFM images. For a few STM measurements, a Pt/Ir tip was used. The STM tips were prepared and formed by controlled indentation into the Au(111) surface. For STM, the tip was grounded, and the bias was applied to the sample during the measurements. All bias voltages mentioned in the manuscript refer to the sample bias. 
The nc-AFM measurements were recorded in frequency modulation mode, operated at a constant amplitude ($A_{p-p}\approx120$~pm).  The amplitude was calibrated  using the normalized time-averaged tunneling current method. \cite{Simon_2007, PhysRevB.81.245322} For the nc-AFM experiments, an HF2LI phase-locked loop from Zurich Instruments was used. The tip was grounded during the nc-AFM measurements. The STM and nc-AFM data were processed using the WSxM software.\cite{Horcas07} \\

\noindent \textbf{LEED}. The LEED patterns of \FeBr \ were acquired using a commercial SpectaLEED from Scienta Omicron, while the \CoBr \ samples were measured with an MCP-LEED from OCI Vacuum Microengineering Inc. The LEED patterns are simulated with LEEDPat4.2 software.\cite{LEEDPAT}\\

\noindent \textbf{Sample preparation}. The Au(111) single crystal (MaTeck) was cleaned by several cycles of Argon ion sputtering followed by annealing to 650~K for 15~min. All sample temperature values provided in this manuscript refer to measurements at a thermocouple that is located in close proximity to the sample. The \FeBr \ (Iron(II) bromide, anhydrous, purity 98\%, Alfa Aesar), FeBr$_{3}$ \ (Iron(III) bromide, anhydrous, purity $\geq$98\%, Alfa Aesar) and \CoBr \ (Cobalt(II) bromide, anhydrous, purity $\geq$97\%, Alfa Aesar) powders were  evaporated in UHV from a Knudsen cell (Kentax GmbH) located in the preparation chamber. The sublimation from a quartz crucible occurred at 550~K (\FeBr/FeBr$_{3}$) and 590~K (\CoBr) at an evaporation pressure of around 10$^{-9}$ mbar. During deposition, the Au(111) sample was kept at 390~K-450~K, the temperatures are indicated in the respective figure captions. The deposition rate of \FeBr, FeBr$_{3}$ and \CoBr \ was checked by a quartz microbalance and the powders were thoroughly degassed respectively before the sublimation to Au(111). The deposition rate of \FeBr, FeBr$_{3}$ and \CoBr \ used in this work are 0.02-0.07~ML/min, respectively. The coverage of the \FeBr\ film was controlled by varying the evaporation time.\\

\noindent \noindent \textbf{Calculations}. The first principles calculations reported here are performed within the framework of the density functional theory (DFT), employing the VASP code.\cite{Kresse1996,Kresse1996b} In the periodic VASP calculations, the projector augmented wave (PAW)\cite{Blochl1994} method was employed to describe the core electrons. The exchange-correlation energy and potential are treated within the spin-polarized generalized gradient approximation (GGA) using  the exchange-correlation functional of Perdew-Burke-Ernzerhof (PBE)\cite{Perdew96} with the DFT-D3 dispersive corrections\cite{Grimme10} (using Becke-Johnson damping). The energy cutoff for the plane wave is kept at 400~eV for the ground state calculations. Energies were converged to 10$^{-5}$~eV and geometries were relaxed until the forces on all atoms were below 0.001~eV/\AA, respectively. A Methfessel-Paxton-smearing \cite{Methfessel1989} of first order with smearing parameter $\sigma$=0.2~eV was used. To model the \FeBr \ structures \ in the gas phase and on the substrate, we have constructed a superstructure corresponding to a $\vec{v}_\mathrm{FeBr_2}=(3,1)$ on $\vec{v}_\mathrm{Au}=(4,1)$ domain. The pristine \FeBr \ structure contains 21 atoms (7 Fe and 14 Br atoms). The gold slab consists of six layers, with the bottom three layers fixed to the bulk geometry and the remaining three layers free to relax. To prevent interactions between the slab and its periodic images and to account for the finite size of the slab model, gas phase systems were computed with 16~\AA \ of vacuum space and surface calculations were computed with 30~\AA \ of vacuum space in the $z$-direction. A set of (6$\times$6$\times$1) $\Gamma$ centered k-point sampling are used for both gas phase and surface calculations. \newline
The formation energy E$_\mathrm{for}$ of defects is calculated using the equation E$_\mathrm{for}$=E$_\mathrm{defective}$ - E$_\mathrm{FeBr_2}$ + $\sum n_i \mu$, \cite{Cayhan21} where E$_\mathrm{defective}$ and E$_\mathrm{FeBr_2}$  represent the total energy of the defective and pristine FeBr$_2$ single layers, and n$_i$ and $\mu$ are the number and chemical potential of the removed atom, respectively. To calculate $\mu$, we use the relation $\mu$$_\mathrm{FeBr_2} =\mu$$_\mathrm{Fe}$ + 2$\mu$$_\mathrm{Br}$, where $\mu$$_\mathrm{Fe}$, $\mu$$_\mathrm{Br}$, and $\mu$$_\mathrm{FeBr_2}$ are the total energies of Fe, Br, and single-layer \FeBr. For calculating  $\mu$$_\mathrm{Fe}$, we assume the stable bulk form of Fe(bcc-Fe) at the Fe-rich limit.\cite{Kulish2017} Hence, $\mu$$_\mathrm{Br}$ can be evaluated from Fe-bulk and the total energy of single-layer \FeBr. \newline
Constant-height and constant-current STM images were simulated within the Tersoff–Hamann model.\cite{Tersoff83,Tersoff85} The tip was placed  $\sim 2$~\AA \ over the plane of the top Br layer for constant-height images at different bias voltages. For the constant-current STM images, the isodensity values are adjusted from 10$^{-3}$ to  114 electron/\AA$^3$. Constant-height nc-AFM simulations were based on the probe particle model established by Hapala, et al. \cite{Hapala2014} and widely used to model the nc-AFM imaging process with functionalized tips.  We assumed a Br at the tip apex as the tip was intentionally crashed softly into the \FeBr \  monolayer. Frequency shift images were calculated for the Br-functionalized tip assuming a harmonic spring stiffness of 0.5 N/m and an effective charge of $-0.03e$ for the probe particle.

\subsection{Acknowledgements}
This work was funded by the German Research Foundation (DFG) through the SFB 953 \textit{Synthetic Carbon Allotropes} (project number 182849149) and the Interdisciplinary Center for Molecular Materials (ICMM) at the Friedrich-Alexander-Universität Erlangen-Nürnberg. We thank Andreas Dörr, Dengyuan Li and Sajjan Mohammad for their experimental support and discussions.  

\subsection{Notes}
We became aware that in a recently published preprint, similar STM measurements of \FeBr \ on Au(111) were presented.\cite{hadjadj2023} However, a different conclusion concerning the structure and chemical composition was reached.

\vspace{1cm}




\bibliography{references}

\clearpage

\end{document}




\onehalfspacing
\small
\clearpage

\tableofcontents

\clearpage

\section{Structural models of 2D transition metal halides}\label{sec:structure}

\begin{figure*}[h!]
\centering
 \includegraphics[width=0.6\paperwidth]{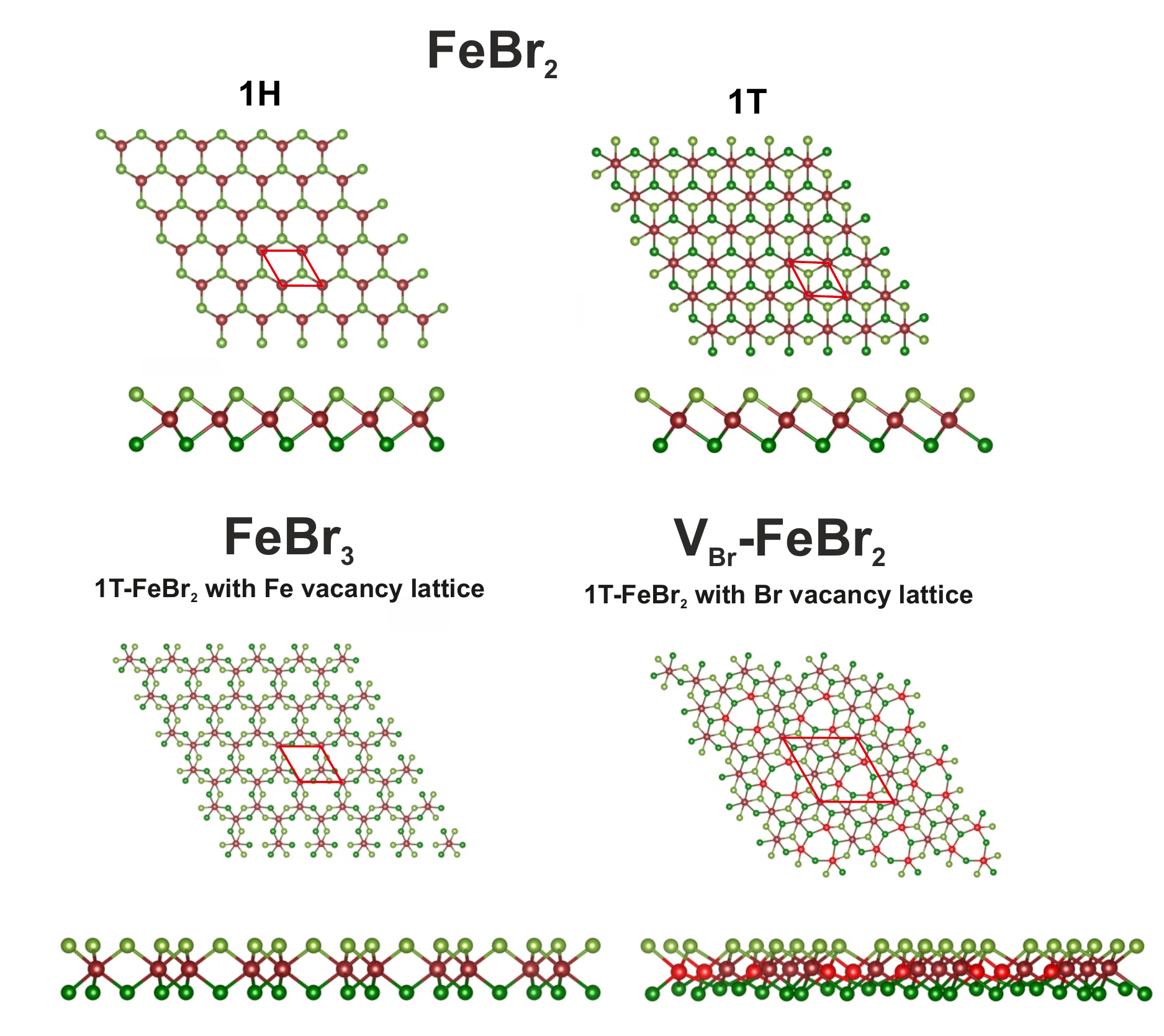}
 \caption{\footnotesize \textbf{Schematic view of the structure of \FeBr and FeBr$_{3}$ monolayers and their polytypes.} Transition metal halides can adopt both trigonal prismatic (1H) or octahedral coordination (1T). However, for FeBr$_{2}$ and CoBr$_{2}$  1T stacking is energetically preferred.\cite{Kulish2017} FeBr$_{3}$ is formed upon incorporating regular Fe vacancies into \FeBr monolayers. Conversely, periodic assemblies of Br vacancies yield the novel \vFeBr films discussed here. Color code: light green, top Br; dark green, bottom Br; dark red, 6-fold coordinated Fe; red, 5-fold coordinated Fe. }  
 \label{fig:FeBr_structural_model}
\end{figure*}

\clearpage

\section{Imaging of the \FeBr superstructure by STM and nc-AFM}\label{sec:imaging}
\begin{figure*}[h!]
\centering
 \includegraphics[width=0.6\paperwidth]{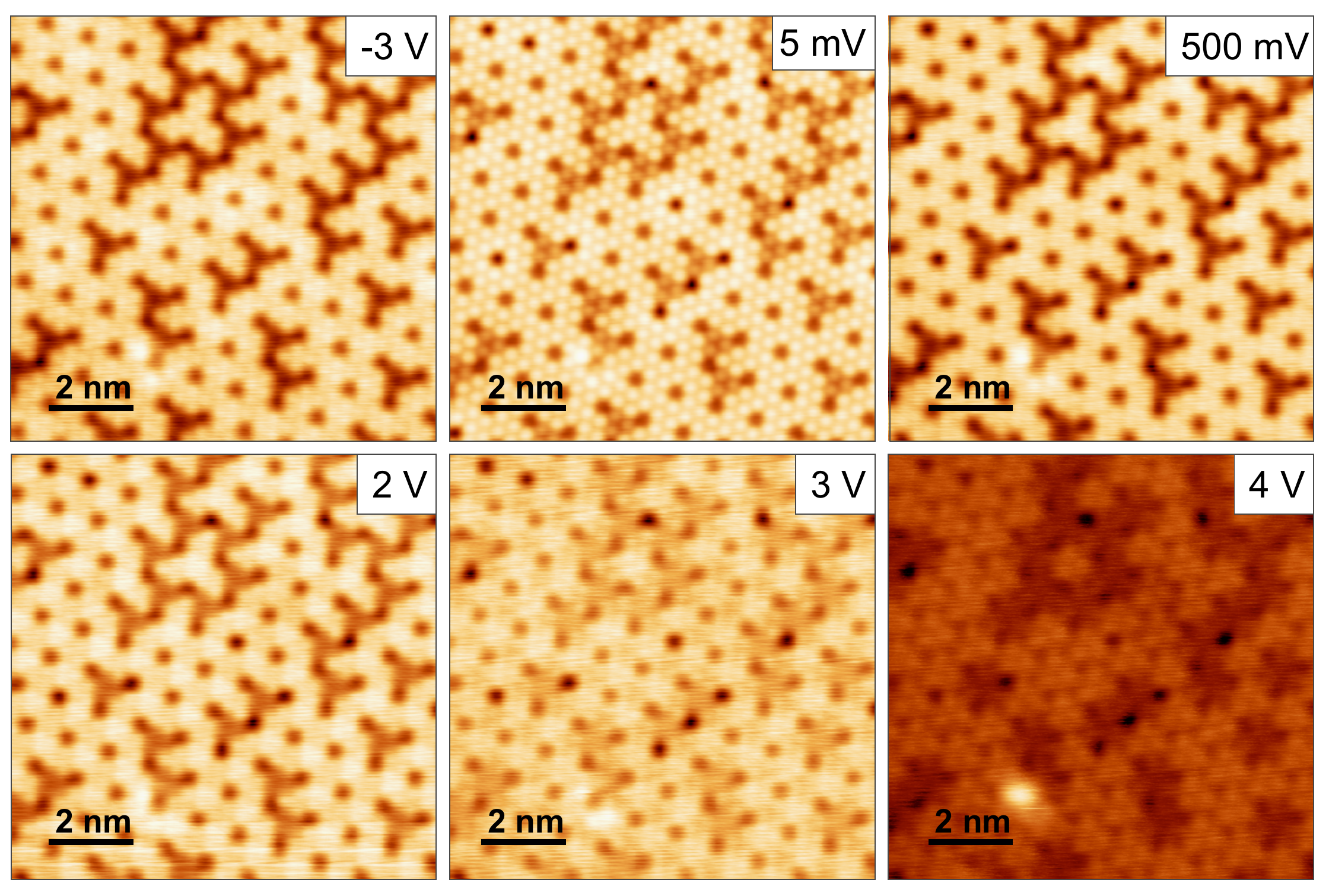}
 \caption{\footnotesize \textbf{Bias-dependent STM images (\textit{I}=560~pA) of the \FeBr structures.}
 The \FeBr powder was deposited on Au(111) kept at 390 K, coverage: 0.5~ML.}  
 \label{fig:FeBr_bias_dependence}
\end{figure*}

\begin{figure*}[h!]
\centering
  \includegraphics[width=0.5\paperwidth]{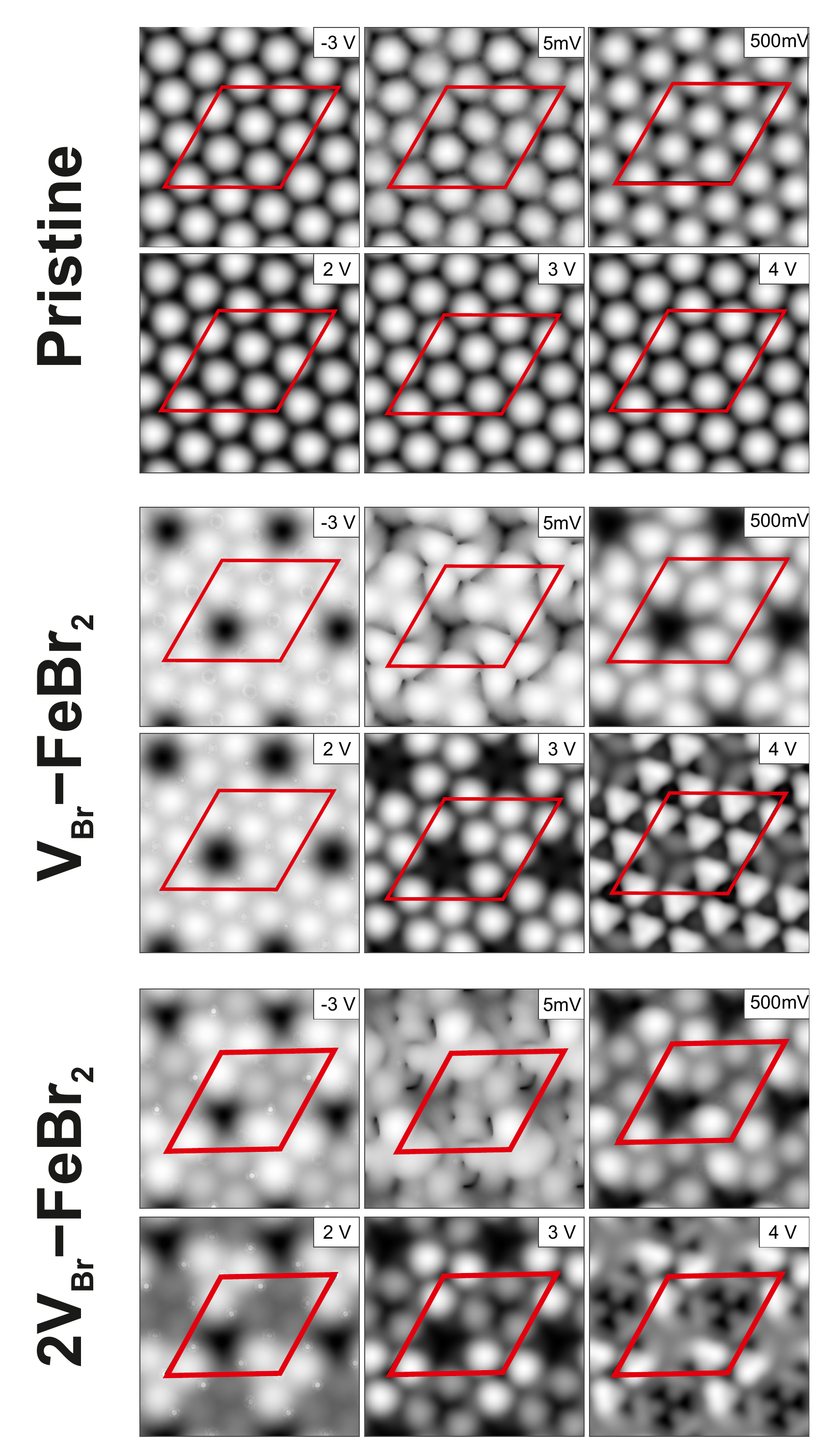}
 \caption{\footnotesize \textbf{Simulated constant-current STM images from pristine and defective \FeBr layers on Au(111).} The defective models include one Br vacancy in the top halide layer as well as one top and one bottom Br defect.}
 \label{fig:FeBr_bias_dependence_calc}
\end{figure*}

The periodic depressions forming the observed superstructure can be explained by either an ordered vacancy lattice in the top halide layer of the transition metal dihalide (TMD) or a \moire pattern resulting from the \FeBr monolayer's rotation on the Au(111) lattice. In a commensurate case, the \moire pattern has a hexagonal unit cell due to the hexagonal symmetry of the Au(111) and \FeBr lattice. We note that the topographic contrast of the superstructure is nearly bias-independent over a large voltage range (-3V to 2V) in constant-current STM, see Fig.~\ref{fig:FeBr_bias_dependence}. Calculated constant-current STM images for the different \FeBr structures feature similar contrast, see Fig.~\ref{fig:FeBr_bias_dependence_calc}. Hence, bias-dependent STM experiments and corresponding DFT calculation give the indication that the superstructure is likely not related to an electronic \moire pattern in agreement with the nc-AFM experiments \\

\clearpage

\begin{figure*}[h!]
\centering
  \includegraphics[width=0.6\paperwidth]{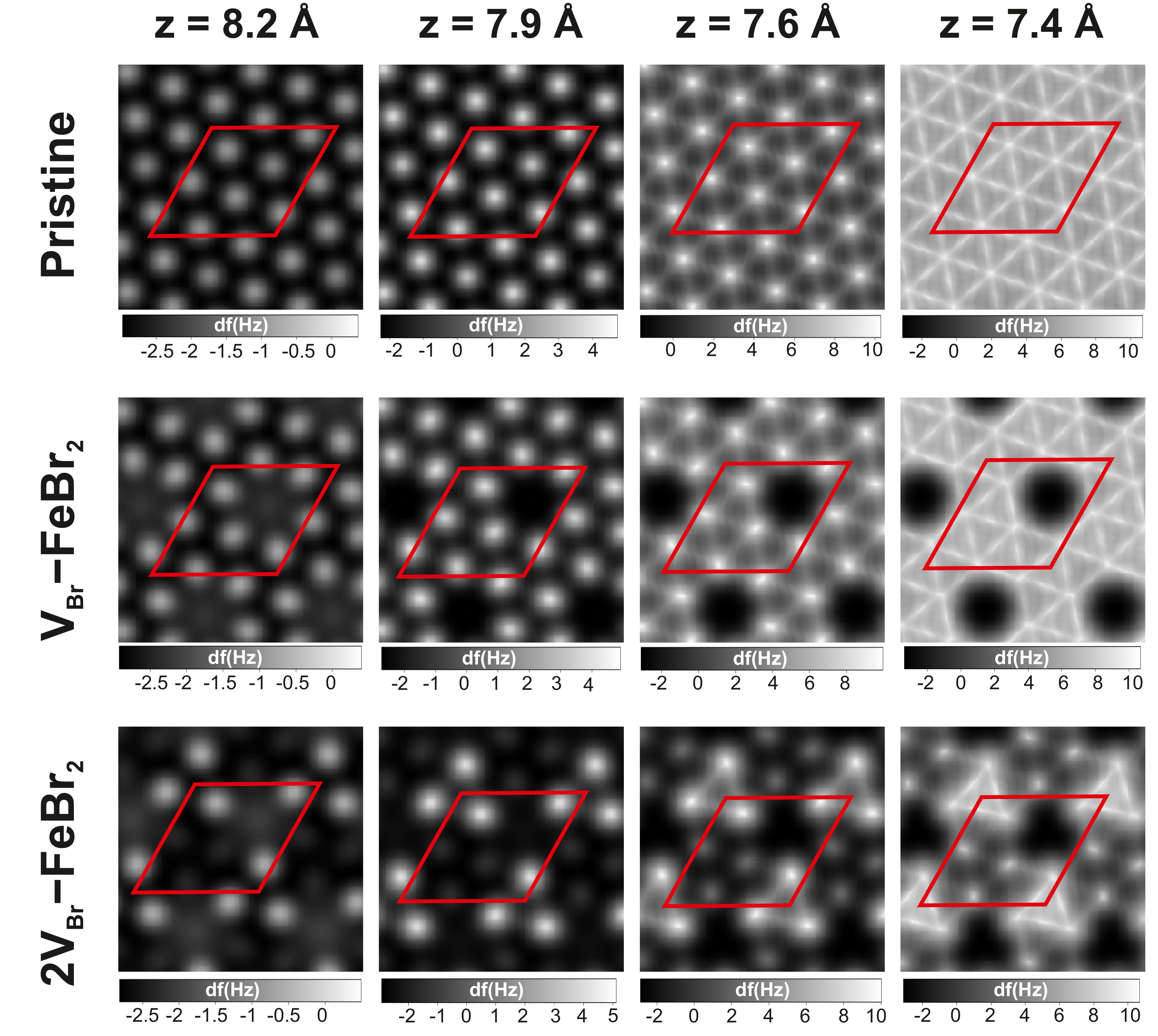}
 \caption{\footnotesize \textbf{Simulated nc-AFM images for pristine and defective \FeBr \ layers on Au(111).}  The nc-AFM simulations were based on the DFT-optimized \FeBr \ geometries on Au(111) using the probe particle model established by Hapala et al. \cite{Hapala2014}.  We assumed a Br at the tip apex as the tip was intentionally crashed softly into the \FeBr \  monolayer. Frequency shift images were calculated for the Br-functionalized tip assuming a harmonic spring stiffness of $0.5$~N/m and an effective charge of $-0.03e$ for the probe particle. } \label{fig:FeBr_bias_dft_ncafm}
\end{figure*}
\clearpage

\begin{figure*}[h!]
\centering
  \includegraphics[width=0.6\paperwidth]{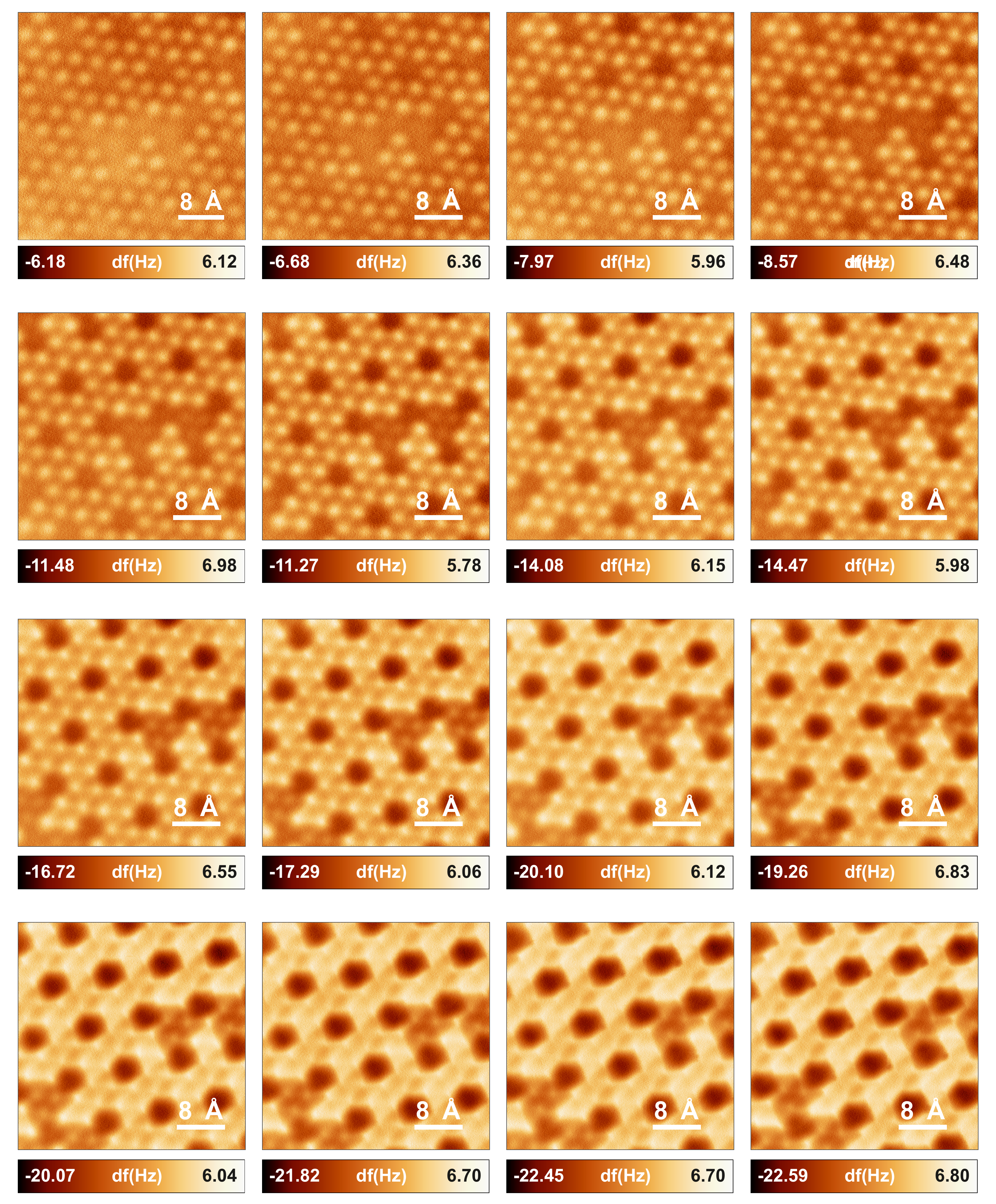}
 \caption{\footnotesize \textbf{Constant-height nc-AFM images of \vFeBr on Au(111) at various tip-sample distances.} The tip-sample distance gradually decreases from top to bottom.}   
 \label{fig:FeBr_ncafm}
\end{figure*}
\clearpage

\section{Calculation of structural parameters of the \vFeBr superstructure}

We rationalize the \vFeBr superstructure based on  the superposition of two rotated commensurate hexagonal lattices. We define the superstructure vector $L=a_\mathrm{Au}\sqrt{m^2+n^2-mn}=a_\mathrm{FeBr_2}\sqrt{r^2+s^2-rs},$ with $\vec{v}_\mathrm{FeBr_2}=(r,s)$ being unit cell vectors of the \FeBr lattice and $\vec{v}_\mathrm{Au}=(m,n)$ being unit cell vectors of the Au lattice, respectively. The two basis vectors spanning the unit cell of the substrate and FeBr$_{2}$, respectively, include angles of 120$^{\circ}$.  In this case, the rotational angles of the different periodic structures are given by \cite{Zeller2017} 
$$\cos(\Phi_\mathrm{vac-lattice,Au})=\frac{m-\frac{n}{2}}{\sqrt{m^2+n^2-mn}}, $$
$$\cos(\Phi_\mathrm{vac-lattice,FeBr_2})=\frac{r-\frac{s}{2}}{\sqrt{r^2+s^2-rs}}, $$\\
where $\Phi_\mathrm{vac-lattice,Au}$ denotes the rotation angle of the vacancy lattice with respect to the Au lattice and $\Phi_\mathrm{vac-lattice,FeBr_2}$ with respect to the \FeBr lattice. \\

\begin{figure*}[h!]
\centering
 \includegraphics[width=0.6\paperwidth]{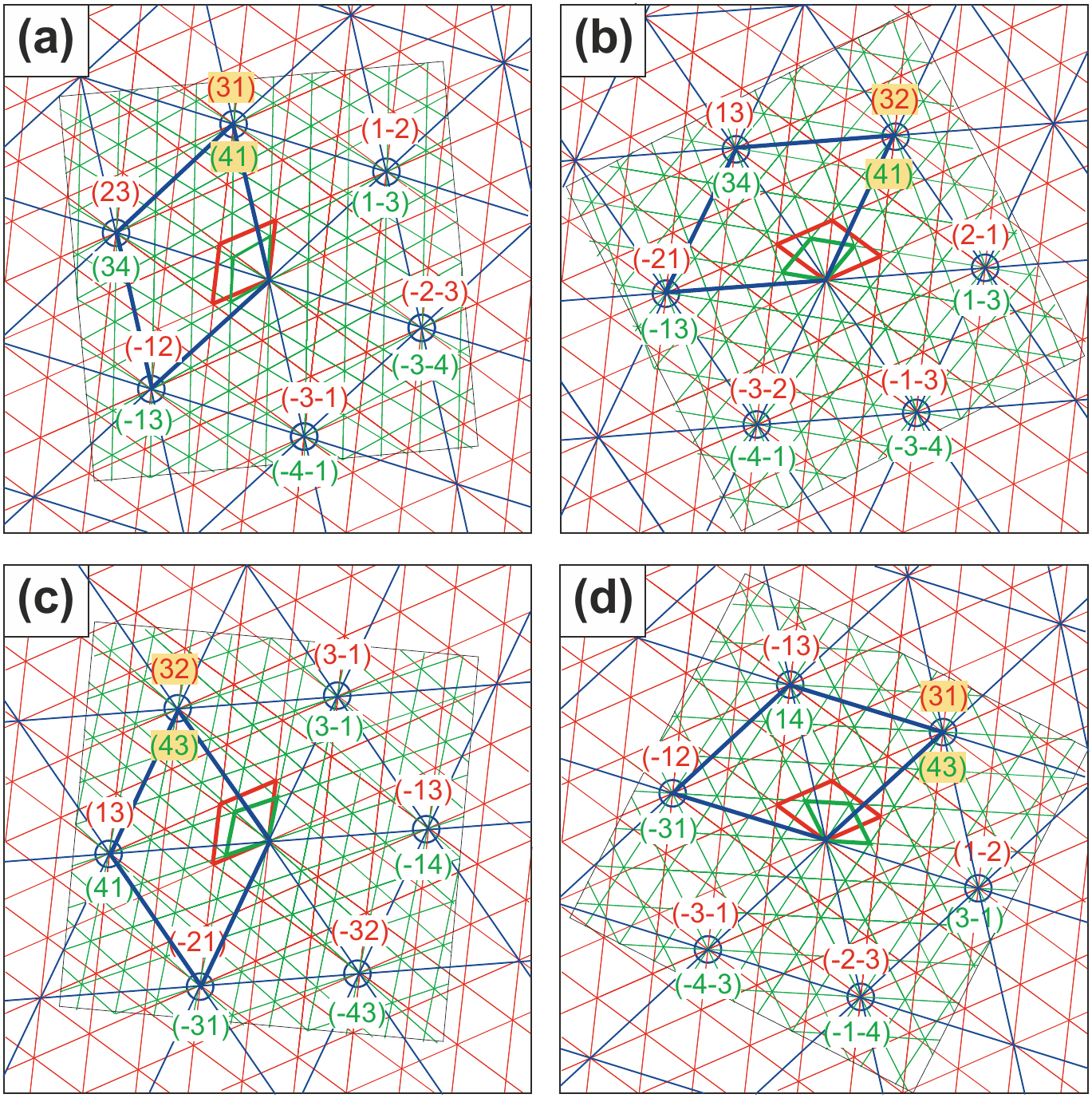}
 \caption[]{\footnotesize \textbf{Schematic representation of all possible domains of \vFeBr.} The \FeBr lattice is plotted in red, the Au lattice in green, and the resulting superstructure pattern in blue. The nomenclature of the vectors $L$ is based on Ref.~\citenum{Guenther2021} and includes angles of $120^{\circ}$ between the basis vectors. The red and green labels express the vectors $L$ rotated by multiples of $60^{\circ}$ in the unit vectors of \FeBr and of the Au substrate, respectively. The angles are given in Tab.~\ref{tbl:unit_cells_powder_SI} for the structures highlighted in yellow.}  
 \label{fig:moirelattices}
\end{figure*}

Table~\ref{tbl:unit_cells_powder_SI} gives an overview of the angles of the experimentally observed and relevant calculated superstructure patterns. There are 4 distinct domains with $L=\sqrt{13}a_\mathrm{Au}$: $\vec{v}_\mathrm{FeBr_2}=(3,1)$ on $\vec{v}_\mathrm{Au}=(4,1)$,  $\vec{v}_\mathrm{FeBr_2}=(3,2)$ on $\vec{v}_\mathrm{Au}=(4,3)$, $\vec{v}_\mathrm{FeBr_2}=(3,1)$ on $\vec{v}_\mathrm{Au}=(4,3)$, and $\vec{v}_\mathrm{FeBr_2}=(3,2)$ on $\vec{v}_\mathrm{Au}=(4,1)$. These results in total in 24 domains, including the ones rotated by multiples of 60$^{\circ}$ with respect to $L$ as depicted in Fig.~\ref{fig:moirelattices}. Thereby, the superstructures in Fig.~\ref{fig:moirelattices}(a)-(b) are described by the matrix $\big(\begin{smallmatrix}
  4 & 1\\
  -1 & 3
\end{smallmatrix}\big)$ and in Fig.~\ref{fig:moirelattices}(c)-(d)
$\big(\begin{smallmatrix}
  4 & 3\\
  -3 & 1\end{smallmatrix}\big)$ with respect to the Au substrate.

 \begin{table*}[h!]
   \footnotesize
     \centering
     \begin{tabular}{l|ccc}
     \hline
     \rowcolor[gray]{.8}
          & \textbf{$\Phi_\mathrm{vac-lattice,FeBr_2}$}& \textbf{$\Phi_\mathrm{vac-lattice,Au}$}& $\Theta_\mathrm{FeBr_2,Au}$ \\  
       \hline
       \textbf{STM/LEED experiment}  & & &\\
       \FeBr powder      & 19.2$\pm1^{\circ}$ / 40.9$\pm1^{\circ}$  & 13.9$\pm1^{\circ}$ / 46.1$\pm1^{\circ}$&  -5.2$\pm1^{\circ}$ / 27.0$\pm1^{\circ}$ \\
       \hline
       \textbf{Calculated structures}  & & &\\
         $\vec{v}_\mathrm{FeBr_2}=(3,1),\vec{v}_\mathrm{Au}=(4,1)$ &   19.11$^{\circ}$&  13.90$^{\circ}$ &    -5.21$^{\circ}$  \\      
	  $\vec{v}_\mathrm{FeBr_2}=(3,2),\vec{v}_\mathrm{Au}=(4,3)$ &   40.89$^{\circ}$&  46.10$^{\circ}$ &   5.21$^{\circ}$  \\      
        $\vec{v}_\mathrm{FeBr_2}=(3,1),\vec{v}_\mathrm{Au}=(4,3)$ &   19.11$^{\circ}$&  46.10$^{\circ}$ &   26.99$^{\circ}$  \\     
         $\vec{v}_\mathrm{FeBr_2}=(3,2),\vec{v}_\mathrm{Au}=(4,1)$ &   40.89$^{\circ}$&  13.90$^{\circ}$ &   -26.99$^{\circ}$  \\                     
        \hline	
    \end{tabular}
     \caption{\footnotesize \textbf{Experimentally measured and calculated rotation angles of \vFeBr \ grown on Au(111)} The additional domains obtained by rotation of multiples of 60$^{\circ}$ with respect to $L$ are outlined in Fig.~\ref{fig:moirelattices}. We note the rotation angles are the same for CoBr$_{2}$ as only the superstructure and not the lattice constants are relevant.}
     \label{tbl:unit_cells_powder_SI}
   \end{table*}

\clearpage
\section{Structure of the \vFeBr superstructure from STM and LEED}

The Br-vacancy superstructure denotes a ($\sqrt{7}\times\sqrt{7})\mathrm{R}19.1^{\circ}$ structure with respect to the \FeBr \ lattice. Hence, we expect a 13.9$^{\circ}$ and 46.1$^{\circ}$ rotation of the superstructure toward the high symmetry axes of the Au lattice, as well as a  $\pm 5.21^{\circ}$ and $\pm 26.99^{\circ}$ rotation of the \FeBr \ lattice toward the Au lattice, see Tab.~\ref{tbl:unit_cells_powder_SI}. 

\begin{figure*}[h!]
 \centering
 \includegraphics[width=0.6\paperwidth]{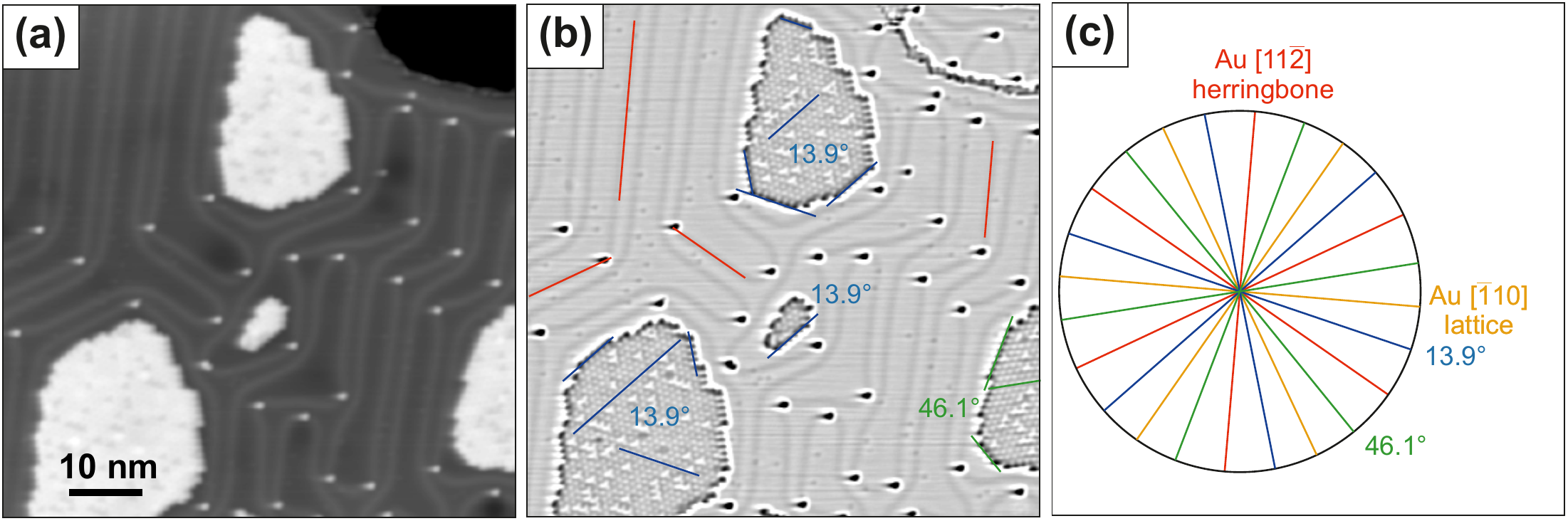}
 \caption{\footnotesize \textbf{Rotational domains in \vFeBr on Au(111) in STM.} The observed $(\sqrt{7}\times\sqrt{7})\mathrm{R}19.1^{\circ}$ superstructure adopts two different orientations towards the Au substrate, i.e. 13.9$^{\circ}$ and 46.10$^{\circ}$ of the superstructure lattice towards the high symmetry axes of Au lattice. (a) STM topography image and corresponding Laplace filtered image in (b). STM parameter: -1 V, 100 pA.}
 \label{fig:FFTpowder}
\end{figure*}

\begin{figure*}[h!]
 \centering
 \includegraphics[width=0.6\paperwidth]{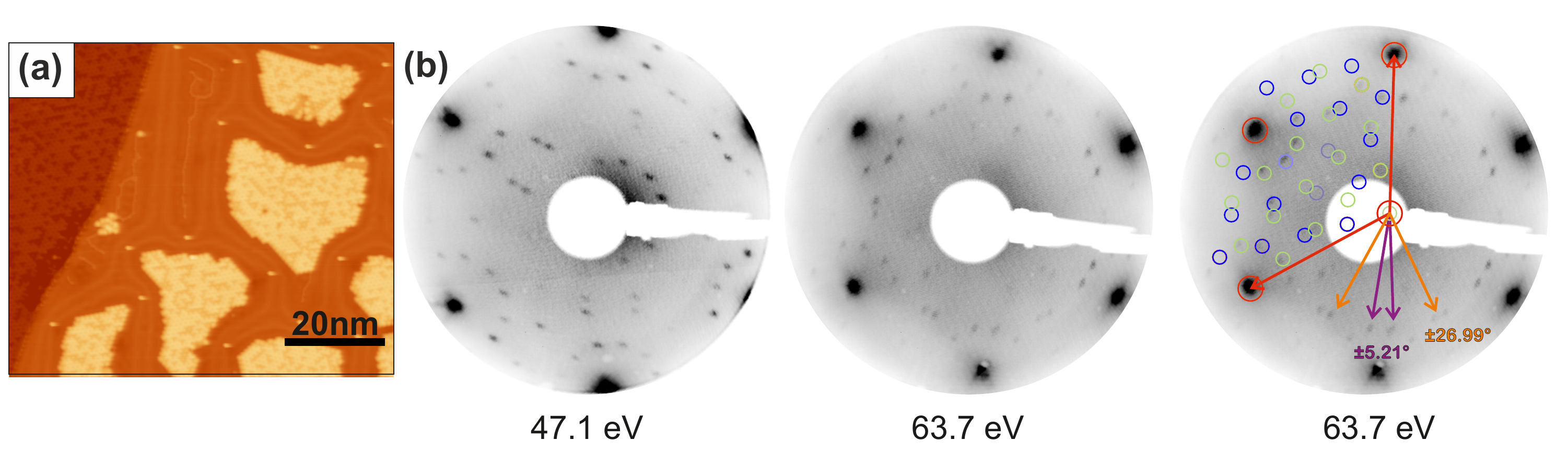}
 \caption{\footnotesize \textbf{LEED of \vFeBr on Au(111) at submonolayer coverage.} (a) STM image of $\approx0.5$~ML on \vFeBr  deposited at 180°C on Au(111) with corresponding LEED measurements at various energies in (b). STM parameter: 100mV, 300 pA. }
 \label{fig:SI_LEED}
\end{figure*}

STM measurements of submonolayer of \FeBr on Au(111) in Fig.~\ref{fig:FFTpowder} confirm that the lattice of the Br-vacancy superstructure has two rotational domains oriented 13.9$^{\circ}$ and 46.10$^{\circ}$ toward the high symmetry axes of the Au lattice. The low-energy electron diffraction (LEED) measurements at different beam energies for submonolayer coverage of \FeBr on Au(111) in Fig.~\ref{fig:SI_LEED} reveal the 24 domains of the \vFeBr structure  with a  $\pm 5.21^{\circ}$ and $\pm 26.99^{\circ}$ rotation of the \FeBr \ lattice toward the Au lattice.  \\

\section{Magnetic ordering in 1T-\FeBr monolayers}

\begin{figure}
 \includegraphics[width=0.8\columnwidth]{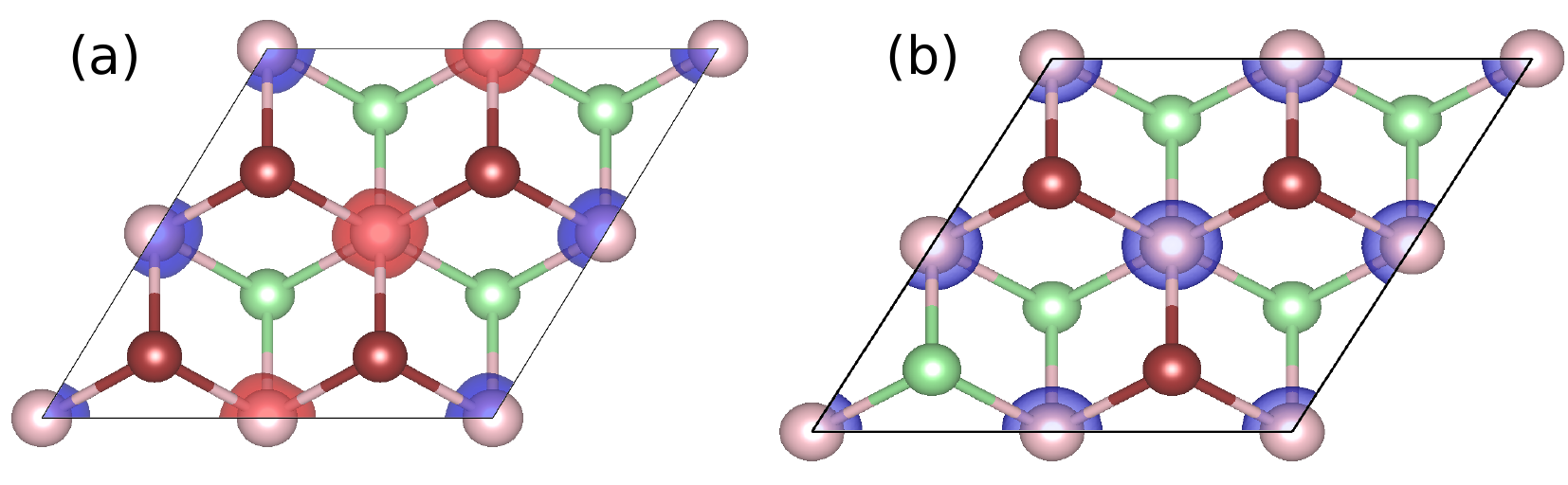}
 \caption{\footnotesize \textbf{ Spin density isosurface plots of \FeBr monolayer for isosurface level of $\pm$0.06~{\AA}$^{-3}$  are displayed, (a) AFM and (b) FM spin arrangement.} Color code for the atoms: red, top Br layer; green, bottom Br layer; pink, Fe atoms.  Color code for the isosurfaces: Red for negative isosurfaces, blue for positive ones.}
 \label{fig:spin-density}
\end{figure}

In order to determine the magnetic state of pristine 1T-\FeBr monolayers, we calculated the energy difference $\Delta$ E between several antiferromagnetic (AFM) arrangements and the ferromagnetic (FM) arrangement of the spins of the iron atoms by subtracting the corresponding total energies E$_{AFM}$ and E$_{FM}$, respectively, i.e., by $\Delta E =E_{AFM}-E_{FM}$. One of the considered AFM spin arrangements and the FM spin arrangement are depicted in Fig.~\ref{fig:spin-density}(a)-(b). The FM state is found to be lower in energy than all considered AFM states. The energy difference between the two arrangements depicted in Fig. S7 equals $\Delta$ E = 78 meV. These results are in accordance with the previous results reported on FeBr$_{2}$.\cite{Kulish2017}.

\clearpage


\section{Additional STM figures}


\begin{figure}[h!]
\centering
 \includegraphics[width=0.6\columnwidth]{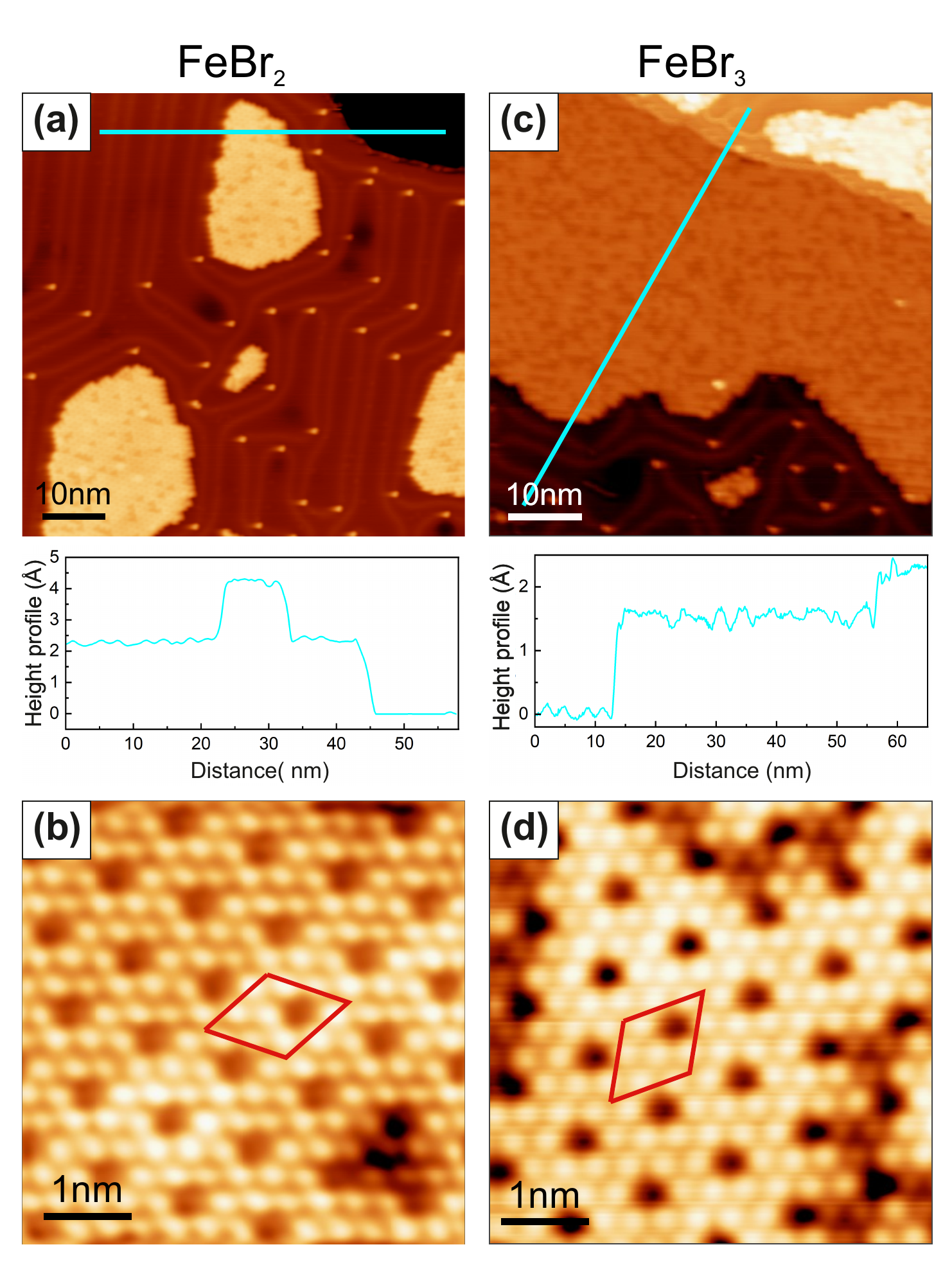}
 \caption{\footnotesize \textbf{Comparison of \vFeBr monolayers on Au(111) using \FeBr and  FeBr$_{3}$ powder.} Overview and detailed STM images of monolayer-thin \vFeBr islands grown from (a-b) \FeBr and (c-d) FeBr$_{3}$  powder deposited on a Au(111) sample, which was kept at 450~K and 390 K, respectively. The coverages of (a-b) are 0.3 ML, 0.4 ML for (c-d).  STM parameters: (a) \textit{I} = 100~pA, \textit{U}= -1~V; (b) \textit{I} = 100~pA, \textit{U}= -5~mV; (c) \textit{I} = 100~pA, \textit{U}= -500~mV; (d) \textit{I} = 100~pA, \textit{U}= -50~mV.}
 \label{fig:FeBr2_3}
\end{figure}

\clearpage


\begin{figure}[h!]
\centering
 \includegraphics[width=1\columnwidth]{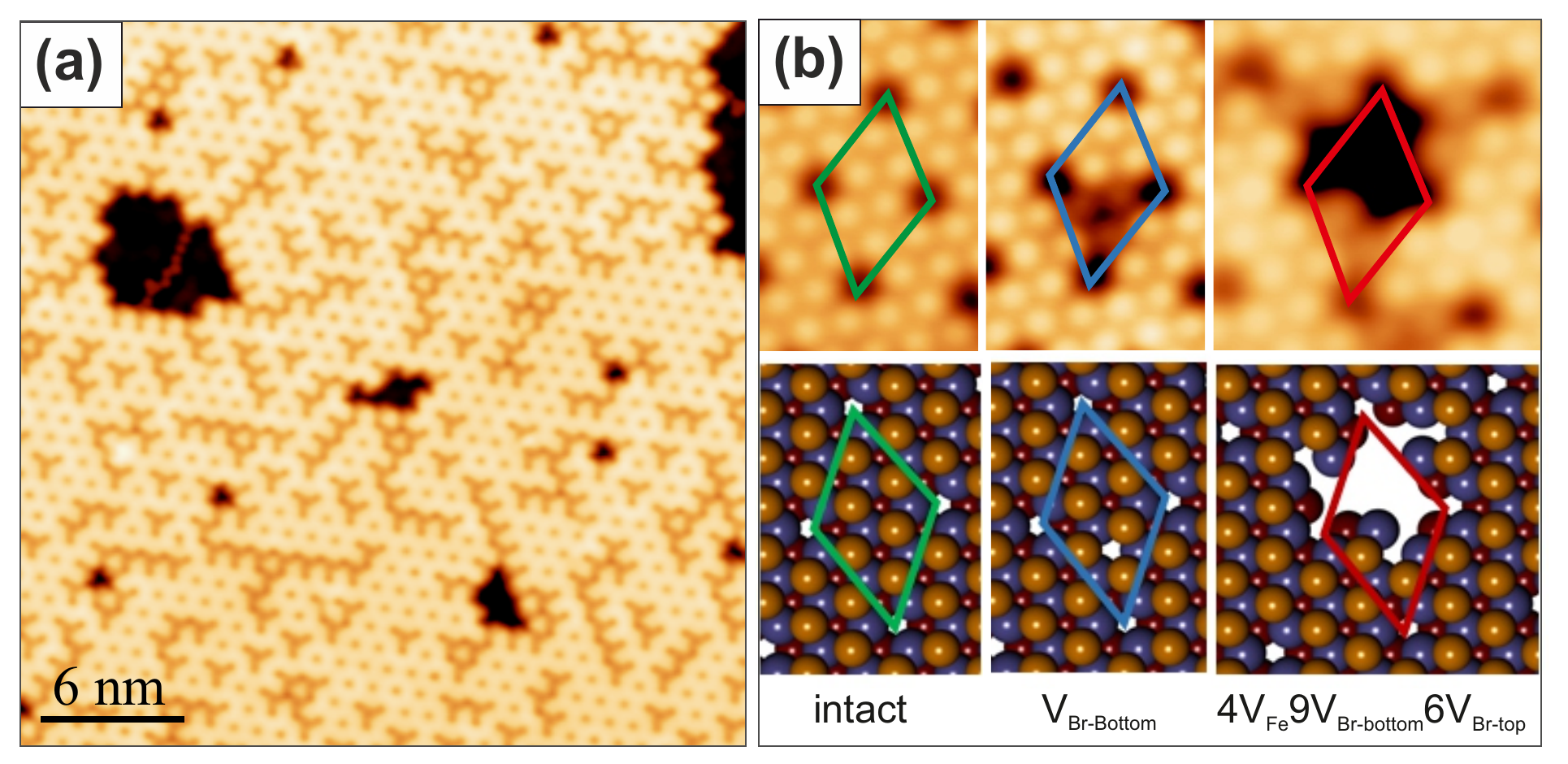}
 \caption{\footnotesize \textbf{STM images of defects in \FeBr on Au(111).} (a) Overview and (b) detailed STM images of defective FeBr$_{2}$. The most frequently observed defects are Br vacancies in the bottom layer of TMD,  located in the bottom part of the blue unit cell. In the upper half of the unit cell a Fe-based vacancy is observed, see red unit cell. While the Br vacancy is frequently observed, Fe defects are very rare and are not observed as single Fe vacancies. Corresponding tentative models are provided in the bottom row. STM parameters: (a) $U=100$~mV, $I=100$~pA; $I=100$~pA, $U=50$~mV. }
 \label{fig:FeBr2_defect}
\end{figure}

\begin{figure}[h!]
\centering
 \includegraphics[width=0.8\columnwidth]{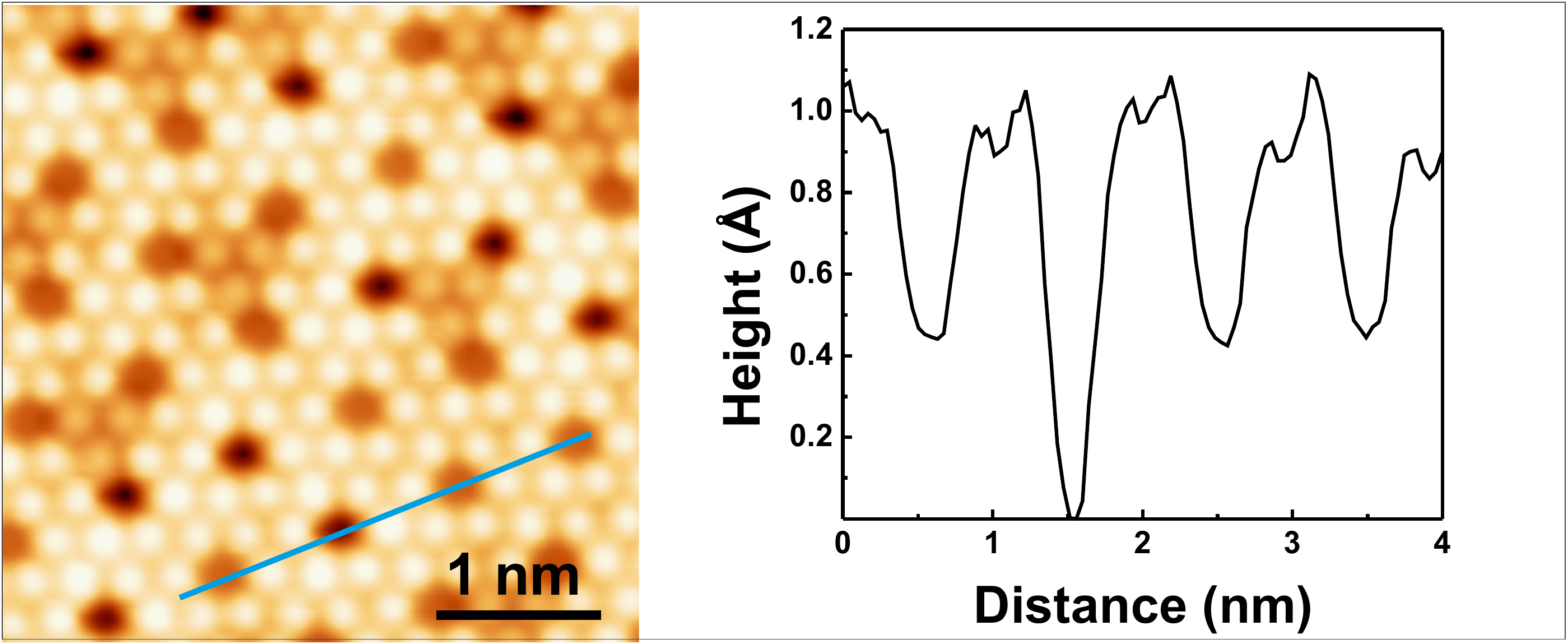}
 \caption{\footnotesize \textbf{Constant-current STM image of \FeBr \ on Au(111) highlighting different imaging contrast of the Br-top vacancies.}  }
 \label{fig:FeBr2_defect_filledpores}
\end{figure}

\clearpage

\bibliography{references}

\clearpage